\documentclass[preprint]{aastex}
\usepackage{amssymb,amsmath,amsfonts,graphicx,epsf,color}
\begin{document}

\title{Temporal Analysis of Dissipative Structures in Magnetohydrodynamic Turbulence}

\author{Vladimir Zhdankin}
\email{zhdankin@wisc.edu}
\affil{Department of Physics, University of Wisconsin-Madison, Madison, WI 53706} 

\author{Dmitri A. Uzdensky}
\email{uzdensky@colorado.edu}
\affil{Center for Integrated Plasma Studies, Physics Department, UCB-390, University of Colorado, Boulder, CO 80309} 

\author{Stanislav Boldyrev}
\email{boldyrev@wisc.edu}
\affil{Department of Physics, University of Wisconsin-Madison, Madison, WI 53706, USA} 

\date{\today}

\begin{abstract}
Energy dissipation is highly intermittent in turbulent plasmas, being localized in coherent structures such as current sheets. The statistical analysis of spatial dissipative structures is an effective approach to studying turbulence. In this paper, we generalize this methodology to investigate four-dimensional spatiotemporal structures, i.e., dissipative processes representing sets of interacting coherent structures, which correspond to flares in astrophysical systems. We develop methods for identifying and characterizing these processes, and then perform a statistical analysis of dissipative processes in numerical simulations of driven magnetohydrodynamic turbulence. We find that processes are often highly complex, long-lived, and weakly asymmetric in time. They exhibit robust power-law probability distributions and scaling relations, including a distribution of dissipated energy with power-law index near $-1.75$, indicating that intense dissipative events dominate the overall energy dissipation. We compare our results with the previously observed statistical properties of solar flares. \\
\end{abstract}

\maketitle

\newpage

\tableofcontents

\section{Introduction}

Intermittency, the inherent inhomogeneity of turbulence, causes energy dissipation to be highly localized in space and in time. Spatial intermittency is manifest by coherent structures, while temporal intermittency is characterized by irregular, bursty events. Both of these phenomena are related, play key roles in the turbulent dynamics, and have important observational consequences. As with any dynamical physics problem, a complete understanding of intermittency can only be claimed when the solution is described in both the spatial and the temporal dimensions.

Intermittency is a challenging theoretical problem that impedes our progress toward a complete theory of turbulence. Many of the methods used to study turbulence, such as energy spectra and correlation functions, are insensitive to intermittency. The methods conventionally employed to measure intermittency include structure functions \citep{muller_etal_2003, chen_etal_2011, podesta_2011}, scale-dependent kurtosis \citep{wan_etal_2012}, topological methods \citep{servidio_etal2009, servidio_etal2010}, and statistics of discontinuities \citep{greco_etal_2009, greco_etal_2009b, zhdankin_etal_2012}; however, these often give incomplete information or are difficult to measure accurately \citep{jimenez_2007, dudokdewit2004, sreenivasan_antonia_1997}. Therefore, in order to better guide theoretical models of turbulence, it is essential to develop new tools for studying intermittency in a robust and informative manner.

The statistical analysis of coherent structures is a possible route forward in this direction. It is convenient, for both practical and theoretical purposes, to treat coherent structures as discrete objects due to their localized nature and their central role in the dynamics. The statistical properties of these structures, including their intensities and morphologies, give insight into the underlying dynamics from which they formed. This information reveals the anisotropy, inhomogeneity, and characteristic scales of the dynamics. Furthermore, such an analysis can be extended beyond simulations and theory, being well suited for a variety of experimental and observational applications, including solar flares, the solar photosphere \citep{cattaneo_1999, bushby_houghton_2005, stein_nordlund_2006}, the interstellar medium \citep{pan_etal_2009, falgarone_etal_2015, boldyrev_etal_2002, kritsuk_etal_2011}, instabilities in fusion devices \citep{carbone_etal2000,antar_etal_2003, dippolito_etal_2004}, and radiative signatures in optically thin astrophysical plasmas, e.g., in black-hole accretion disk coronae \citep{dimatteo_etal_1999}, hot accretion flows \citep{eckart_etal_2009}, jets \citep{albert_etal_2007}, pulsar wind nebulae \cite[e.g.][]{tavani_etal_2011,abdo_etal_2011}, and hot gas in galaxy clusters.

The statistical analysis of spatial structures has been a fruitful approach to studying the intermittency of turbulence in numerical simulations. The methods for identification of structures vary between different studies, but one of the simplest and most common approaches is to identify structures as regions in space bounded by an isosurface of the relevant field. For example, this was applied in hydrodynamic turbulence to study dissipative vorticity filaments \citep{jimenez_etal1993, moisy_jimenez2004, leung_etal2012}, which revealed that the radii of intense filaments scale with the Kolmogorov microscale while lengths occupy large scales. More recently, similar methods were applied to study current sheets and other dissipative structures in simulations of plasma turbulence, notably in 2D magnetohydrodynamic (MHD) turbulence \citep{servidio_etal2009, servidio_etal2010}, 3D MHD turbulence \citep{uritsky_etal2010, zhdankin_etal2013, zhdankin_etal2014}, the kinematic dynamo \citep{wilkin_etal2007}, ambipolar diffusion MHD \citep{momferratos_etal_2014}, boundary-driven MHD \citep{wan_etal_2014}, and decaying kinetic turbulence \citep{makwana_etal_2015}.

These previous studies succeeded in describing the morphology and scaling properties of intermittent structures, but are incomplete in the sense that they give no information on how the structures evolve in time. For this reason, very little can be said about the dynamics, even if one assumes statistical stationarity. Important temporal information about intermittent structures includes their characteristic timescales, stability, motion, interactions, and impulsiveness. If one is interested in understanding these temporal aspects of intermittency, then a broader framework must be developed.

Temporal intermittency was investigated to a limited extent in hydrodynamic turbulence. The temporal statistics of vortices in 2D turbulence \citep{carnevale_etal1991, whitcher_etal2008, pasquero_etal2002} and decaying Charney-isotropic geostrophic turbulence \citep{mcwilliams_etal1999} gave insight into global changes in the population size and morphology of vortices. However, the structures were not entirely treated as temporal objects in these studies, which requires an algorithm for tracking the structures through time. This can be much more difficult to design and implement than the algorithms used to identify and characterize structures in individual fixed-time snapshots \citep{storlie_etal2004, storlie_etal_2009}. To the best of our knowledge, apart from our precursor paper \citep{zhdankin_etal_2015}, no systematic studies of spatiotemporal structures were previously undertaken for hydrodynamic or MHD turbulence in 3D, and only limited studies were performed in 1D and 2D systems \citep[e.g.,][]{aubry_etal1991, daviaud_etal1990, jung_etal2000,colovas_andereck1997}.

The objective of this paper is to extend the framework previously used for the statistical analysis of spatial dissipative structures in numerical simulations of MHD turbulence \citep{zhdankin_etal2013, zhdankin_etal2014} into the temporal realm. We therefore consider the statistical properties of 4D spatiotemporal objects which represent structures evolving through time. These objects are dissipative processes, analogous to flares in astrophysical systems, which, in general, may involve many interacting coherent structures. We consider the distributions and scaling relations for process characteristics including the total dissipated energy, peak energy dissipation rate, duration, and geometric scales; measures of complexity such as the number of interacting structures and types of interactions (mergers and divisions); and the temporal evolution of individual processes. This novel methodology is applied to study intermittency in numerical simulations of driven incompressible MHD turbulence; however, it can also be applied to hydrodynamic turbulence, kinetic plasma turbulence, and other complex dynamical systems exhibiting spatiotemporal self-organization.

The analysis presented in this paper addresses several fundamental questions regarding the intermittency of MHD turbulence. One key question is whether, in the limit of high Reynolds number, the overall energy dissipation is dominated by a few intense, long-lasting events residing at large scales, or by many weak, short-lived events residing near the dissipation scale. Another question is whether there is an inherent relationship between spatial intermittency and temporal intermittency, e.g., whether larger structures better retain their coherency in time. A third question is whether the dissipative events show any characteristic temporal asymmetry, e.g., impulsive onset followed by a slow decay \citep{bhattacharjee_2004}.

The primary questions addressed in this paper for incompressible MHD turbulence are also fundamentally important for the solar corona. In fact, our approach has many similarities with observational studies of solar flares \citep{crosby_etal1993, shimizu1995, boffetta_etal1999, parnell_jupp2000, hannah_etal2008, aschwanden_etal_2000b, uritsky_etal2013, uritsky_etal_2007, veronig_etal_2002, aschwanden_etal_2014} and stellar flares \citep{benz_etal_1994,audard_etal_1999,collura_etal_1988, pallavicini_etal_1990, gudel_etal_2003, telleschi_etal_2005}. In these studies, the time-series of extreme UV, soft X-ray, and hard X-ray emissions from the Sun are used to characterize solar flares. Measured quantities include the size, duration, peak intensity, and fluence of the flares, from which the dissipated energy is inferred. In particular, the probability distribution for dissipated energy is of central importance due to its role in assessing the feasibility of the nanoflare model for coronal heating \citep{parker_1983, parker_1988}. This distribution is observed to obey a power law across eight orders of magnitude, with an index generally close to $-1.8$ \citep{aschwanden_etal_2000b}, although the precise value of the index varies significantly between different studies depending on the time period, region, type of emission, and methods used to identify the flares. Most importantly, this index is shallower than the critical value of $-2$, suggesting that nanoflares do not dominate the overall heating of the solar corona \citep{hudson_1991}.\footnote{Given the distribution for energy dissipation, $P(E) \sim E^{-\alpha}$, the critical index is derived by noting that the total energy dissipation, $E_\text{tot} \propto \int_{E_\text{min}}^{E_\text{max}} E P(E) dE$, scales with the lower bound $E_\text{min}$ if $\alpha > 2$ and with the upper bound $E_\text{max}$ if $\alpha < 2$.}

This analogy with the analysis of solar flares suggests a practical application for the statistical analysis of spatiotemporal structures: to assess whether turbulence plays a role in the intermittent energy dissipation of the solar corona. It is presently unknown whether the complex dynamics of the solar corona arise from self-organized criticality, turbulence, or some other phenomena \citep{georgoulis_2005, uritsky_etal_2007, uritsky_etal2013}. This problem can eventually be addressed by a better comparison between numerical simulations of the various models with observations. Our methodology is ideal for making this comparison. As a first step in this direction, we compare the results in this paper to the observed statistical properties of solar flares, collected from a number of previous studies. We find several nontrivial similarities in both cases, despite the anticipated differences between driven incompressible MHD turbulence and the essentially force-free dynamics of the solar corona. This suggests that MHD turbulence may play a role in the energetics of the corona, a possibility which should be investigated more carefully in future studies.

The outline of this paper is as follows. We first describe the basic concepts and subtleties associated with the temporal analysis of structures in Section~\ref{sec2}. We then outline the technical aspects of our procedure, based on extending our previous framework for the statistical analysis of spatial structures \citep{zhdankin_etal2013, zhdankin_etal2014}, including detailed discussion of the algorithms involved, in Section~\ref{sec3}. In Section~\ref{sec4}, we introduce the measurements that are made on spatiotemporal dissipative structures. The results of our analysis on the combined spatial and temporal intermittency of energy dissipation in 3D MHD turbulence are described in Section~\ref{sec5}. We discuss the implications of our results in Section~\ref{sec6}, which includes a comparison of our results with the observed statistical properties of solar flares. We summarize our conclusions in Section~\ref{sec7}.

\section{Background} \label{sec2}

\subsection{Intermittency in MHD turbulence}

MHD describes the macroscopic dynamics of an electrically conducting fluid, such as a plasma (see, e.g. \cite{biskamp2003}). Given a uniform background magnetic field ${\bf B}_0 = B_0 \hat z$ that is strong relative to turbulent fluctuations (estimated by the rms value, $b_\text{rms}$), and assuming that gradients along the background field are small relative to those in the perpendicular direction, the incompressible MHD equations can be written in a reduced form as
\begin{eqnarray}
\left(\frac{\partial}{\partial t}\mp\boldsymbol{V}_A\cdot\nabla_\parallel\right)\boldsymbol{z}^\pm+\left(\boldsymbol{z}^\mp\cdot\nabla_\perp\right)\boldsymbol{z}^\pm  &=& -\nabla_\perp P 
+\nu\nabla_\perp^2\boldsymbol{z}^\pm +\boldsymbol{f}_\perp^\pm \nonumber \\
\nabla_\perp \cdot \boldsymbol{z}^\pm &=& 0 \, ,
\label{rmhd-elsasser}
\end{eqnarray}  
where $\boldsymbol{z}^\pm=\boldsymbol{v}\pm\boldsymbol{b}$ are the Els\"asser variables (with directions strictly perpendicular to ${\bf B}_0$), $\boldsymbol{v}$ is the fluctuating plasma velocity, $\boldsymbol{b}$ is the fluctuating magnetic field (in units of the Alfv\'en velocity, $\boldsymbol{V}_A={\boldsymbol{
B}}_0/\sqrt{4\pi\rho_0}$, where $\rho_0$ is plasma density), $P$ is the total pressure, $\nabla_\perp$ is the gradient in the $(x,y)$ directions while $\nabla_\parallel$ is the gradient in the $z$ direction, and $\boldsymbol{f}^\pm_\perp$ is the large-scale external forcing. For simplicity, the fluid viscosity $\nu$ is taken to be equal to the magnetic diffusivity $\eta$, and both are assumed to be uniform in space. In the reduced MHD approximation, only the $z$-component of current density $j = j_z = \hat{z} \cdot \nabla_\perp \times \boldsymbol{b}$ and vorticity $\omega = \omega_z = \hat{z} \cdot \nabla_\perp \times \boldsymbol{v}$ are retained; these two scalar fields contain complete information to describe the dynamics as well as energetics.

Energy loss from the system is governed by resistive dissipation and viscous dissipation, with respective energy dissipation rates per unit volume given by $\epsilon_\eta = \eta j^2$ and $\epsilon_\nu = \nu \omega^2$. Since the current density and vorticity are in direct correspondence with energy dissipation rates, they are logical dynamical fields in which to study the intermittency of dissipation. In this work, we focus on resistive dissipation only for simplicity; it is straightforward to apply the methods to obtain similar statistical results for the viscous dissipation.

Intermittency in strong MHD turbulence is characterized by the appearance of thin, quasi-2D ribbon-like structures in the current density, known as (electric) current sheets, with similar structures in the vorticity field. These current sheets are anisotropic in three directions, with the largest scale coinciding with the direction of the local mean field, which is the direction of the guide field to good approximation.  Their lengths and widths are distributed mainly in the inertial range, while the thicknesses are strongly localized inside the dissipation range \citep{zhdankin_etal2013, zhdankin_etal2014}. The current sheets are notable for their role in heating, particle acceleration \citep{kowal_etal_2012}, and magnetic reconnection \citep{biskamp_1986}.

\subsection{General remarks for temporal analysis}

In this paper, we aim to study the temporal properties of intense, spatially-coherent structures produced by intermittency in 3D turbulence. For concreteness, we consider current sheets in MHD turbulence. Despite their prominence, intermittent current sheets occupy a small volume that represents the tail of the probability distribution for current density $j(\boldsymbol{x})$.  To systematically study these structures, we employ a framework based on identifying and characterizing the regions in space with current densities exceeding some fixed threshold value, $j_\text{thr}$, which is the only parameter inherent to the methodology. Each structure is then represented as a 3D volume bounded by an isosurface of $j$.

It is straightforward in principle to extend this procedure into the temporal realm by applying the same threshold criterion to the 4D spatiotemporal field $j(\boldsymbol{x},t)$, thereby obtaining 4D spatiotemporal structures. However, although simple in principle, it is challenging in practice to analyze a high-resolution 4D data set in this way. Assuming that the analysis is applied to post-processed data from well-resolved simulations stored on a computer system, the data storage conditions limit the total number of available snapshots, negatively affecting the possible time cadence of snapshots and size of the overall time interval. Hence, the time resolution of post-processed data must generally be worse than the internal time resolution of the simulation. This is a serious issue since the data must be well-resolved in all dimensions in order to properly resolve and track structures across a wide range of scales.

A related problem is that the primary memory limits the amount of data that can be loaded by the analysis program at any given time, requiring the numerical procedures to work with small pieces of the overall data set at a time. Therefore, a feasible temporal analysis should be based on first identifying the structures in the spatial dimensions of a given snapshot, and then tracking them through time, from their formation to their destruction. The algorithms required to perform this task are rather complex, and must be designed in a robust and efficient manner. In particular, there needs to be an algorithm that accurately associates structures in one snapshot with their time-evolved counterparts in subsequent snapshots. A fundamental challenge here lies in the fact that there may not be a unique correspondence between a structure in one snapshot and a structure in the adjacent snapshot, due to mergers and divisions.

The interactions between structures cause the notion of a coherent time-evolving structure to be ambiguous. Instead, the objects of central importance are the \it processes \rm involving structures. These processes can be variously thought of as sets of interacting structures, as dissipative events, or as flares (if we equate the dissipated energy with outgoing radiation). More generally, they can be thought of as branched spatiotemporal structures.

\subsection{Classification of processes \label{class}}

\begin{figure}
\includegraphics[width=8cm]{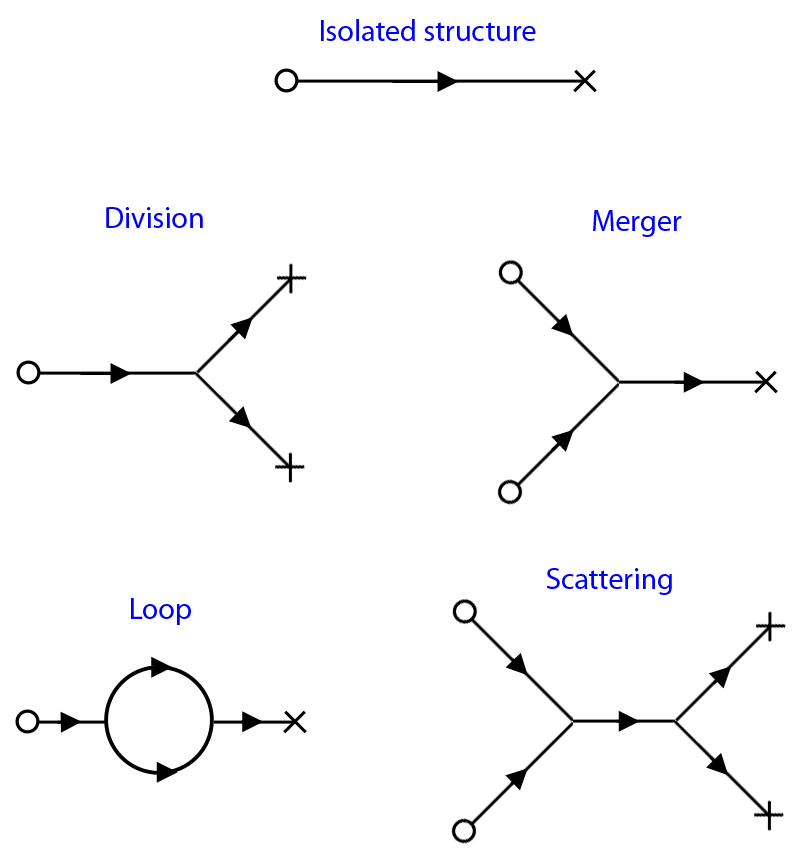}
   \centering
   \caption{\label{fig:interactions} Diagrams of some simple processes, where formation is represented by O, interaction by a vertex, and destruction by X. An isolated structure is a process with no vertices. Division and merger processes are the next simplest case, with a single vertex each. Higher-order processes such as loops and scatterings have a larger number of vertices or vertices with more paths.}
 \end{figure}

In order to build an intuition and facilitate the subsequent discussion of processes, we now describe a convenient classification scheme for processes. This allows processes to be visualized diagrammatically, which has some superficial similarities to Feynman diagrams for quantum mechanical scattering processes \citep{feynman_1948}; however, none of the mathematical symmetries characterized by the Feynman rules carry over, since, to the best of our knowledge, isosurfaces in the current density described by the MHD equations contain no conserved quantities. Regardless, the following classification scheme is a simple way to describe processes and their complexity.

We first introduce some terminology. We define a \it state \rm to be an individual spatial structure at fixed time, which represents the basic building block of processes. We assume that the states are given at times spaced by an infinitesimal increment $dt$. We also assume that there exists a map between all states at any time $t$ to other states at time $t - dt$ and $t + dt$, which represents the instantaneous temporal evolution of structures from one state to another state. We define a \it path segment \rm to be a bijective (i.e., one-to-one) sequence of states under this map, which represents the coherent temporal evolution of an individual structure while it does not interact with any other structures. We then define a \it path \rm to be a path segment with bijectivity breaking down only at the initial and final states in the sequence. A process is then described as a set of paths that are (non-bijectively) connected at their endpoints, which represents a set of interacting structures.

Diagrams of some simple conceivable processes are shown in Fig.~\ref{fig:interactions}. Here, we represent paths schematically as lines with an arrow marking the direction of time. If the path begins by spontaneous formation, i.e., from a peak that grows to exceed the detection threshold, then we mark the beginning of the path with an O. If the path ends by spontaneous destruction, i.e., from a structure that recedes below the detection threshold, then we mark the end of the path with an X. The third possibility is for the path to start or end with an interaction. Interactions between structures are represented by \it vertices \rm connecting sets of three (or more) paths. A process schematically consists of a set of paths connected by a set of vertices.

The simplest process is the evolution of an \it isolated \rm structure, i.e., a structure that is formed and then destroyed without interactions. Isolated structures are described by a single path and have well-defined histories with well-defined properties. Other structures undergo at least one interaction. The two simplest processes involving an interaction are the \it division \rm of one structure into two structures and the \it merger \rm of two structures into one structure. Since an interaction is non-bijective, the structures in these processes do not have completely well-defined histories, so many of the quantities used to describe isolated structures are ambiguous. However, we will see that a meaningful set of more general characteristics can be introduced.

We claim that the most logical approach for a temporal analysis is to study processes rather than individual spatially-coherent structures, i.e., paths, which lose their identity upon interacting. This is also the most conservative approach, as it requires no fundamental changes to the methodology used for the statistical analysis of spatial structures at fixed time, and requires no ad-hoc assumptions to treat the interactions. We also find that the statistical trends are more robust for processes rather than paths.

\section{Methods} \label{sec3}

\subsection{Outline of procedure}

\begin{figure}
\includegraphics[width=16cm]{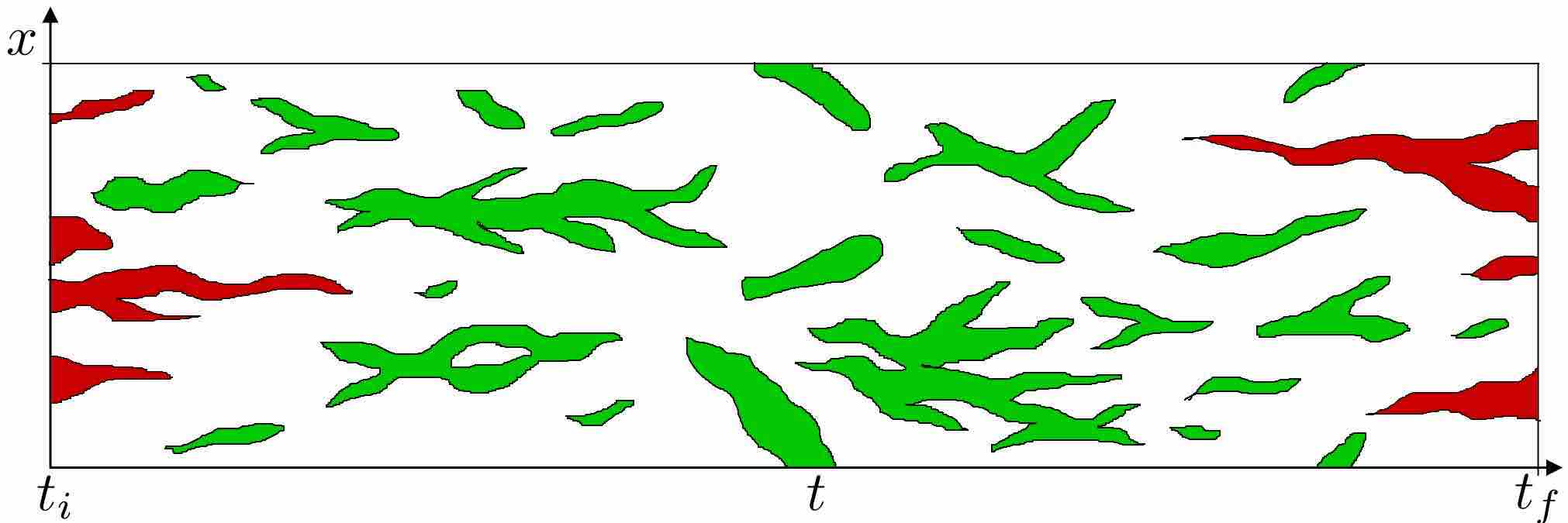}
   \centering
   \caption{\label{fig:scheme} A schematic of structures evolving in time, highlighted in green (shown in one-dimensional space for clarity). Structures from the initial and final states are marked in red.}
 \end{figure}

In this section, we describe our algorithms for the temporal analysis. Since the procedure in its entirety is rather complicated, we include a brief outline in this subsection. A more detailed description is presented in the rest of this section. The procedure rests on a hierarchy of steps: first, we find sets of contiguous points (above the threshold) in each snapshot to obtain the spatial structures; next, we find sets of bijectively-connected states to obtain the paths; finally, we find sets of connected paths to obtain the processes. A schematic of the final result is shown in Fig.~\ref{fig:scheme}, where processes (colored in green) of varying complexity are identified on the space-time lattice.

\begin{enumerate}
\item {\bf Identify all states (i.e., spatial structures) in each snapshot, and represent temporal connectivity by constructing a map between states in adjacent snapshots.}
\begin{enumerate}
\item Load initial snapshot and determine states (by using threshold algorithm).
\begin{itemize}
\item Store constituent points of each state in a temporary array;
\item Perform measurements on states and store in a permanent array;
\end{itemize}

\item For $k = 2, \dots, N_{\text{snap}}$, do the following:
\begin{itemize}
\item Load $i$th snapshot, determine states, and store constituent points and measurements;
\item Construct the evolution map, which associates states in snapshot $k-1$ with temporally-connected states in snapshot $k$, and vice-versa (do this by comparing constituent points of structures in both snapshots); remove temporary array for snapshot $k-1$;
\end{itemize}
\end{enumerate}

\item { \bf Identify paths.}
\begin{enumerate}
\item Obtain paths from states that form a bijective sequence under the evolution maps.
\item Perform measurements on paths by referencing the constituent states.
\item For each path, construct (non-bijective) map that identifies other paths connected to it.
\end{enumerate}

\item { \bf Identify processes.}
\begin{enumerate}
\item Obtain processes from sets of connected paths.
\item Perform measurements on processes by referencing the constituent paths.
\item Treat processes that exist during the initial snapshot or final snapshot as special cases (incomplete processes).
\end{enumerate}
\end{enumerate}

\subsection{Identification of states (spatial structures) \label{sec:vol}}

We first describe our algorithm to identify the states on each snapshot, i.e., the spatial structures at fixed times. We define states as spatially-connected sets (i.e., clusters) of points with current density magnitudes larger than a fixed threshold, $j_\text{thr}$. Two points on the lattice are considered spatially connected if the distance between the points is strictly less than two lattice spacings, i.e., one is contained in the other's 26 nearest neighbors.

We note that this is not the only conceivable way to define a state; one can, for example, use a variable threshold based on local field quantities such as the local peak current density \citep{zhdankin_etal2013}. However, a fixed threshold appears to be the simplest approach, having only one free parameter ($j_\text{thr}$) and no need for ad-hoc treatment of special cases (e.g., mergers). The main drawback of using the fixed threshold is that it generally gives a large number of unresolved structures due to peaks near the threshold; these do not exhibit physically meaningful scaling relations and must be carefully ignored. Fortunately, they are only manifest as noise in the low-intensity, small-scale regime of parameter space and generally have a negligible contribution to the total energy dissipation and volume of the structures. To some extent, unresolved structures are inevitable regardless of how they are defined, since there will always be a population of small, short-lived processes representing structures that barely cross the detection criteria. Filtering procedures can be applied to remove the population of unresolved, noisy structures residing near the threshold, but for simplicity we do not apply any filtering in our analysis.

The definition of a state can affect how a \it{given }\rm process appears; it may determine, for example, whether the close approach of two structures is registered as a merger or not. The threshold value may likewise have a similar effect. However, we would like to emphasize that the \it{statistical }\rm conclusions should be, and in our experience are found to be, broadly consistent regardless of the method or threshold.

Our algorithm identifies states by scanning the lattice for points with current densities above the threshold. For each such point found, the neighboring points satisfying $|j| > j_\text{thr}$ are identified, then neighbors of those neighboring points satisfying $|j| > j_\text{thr}$, and so on, until no more points remain. The coordinates of the constituent points of the state are stored in an array for later use, and then the lattice is scanned for any additional states (while ignoring the points already found to belong to a state). Once every state in a snapshot is identified, measurements are performed on the states (see Section \ref{sec4}) and stored for later use.

\subsection{Temporal association between states in adjacent snapshots}

In this subsection, we consider the connectivity of states in the temporal dimension. We describe our algorithm for identifying the time-evolved counterparts of a state, i.e., the states in the adjacent snapshots that represent the evolved structure. The result is a (non-bijective) map associating the set of states in one snapshot with the set of states in an adjacent snapshot, representing the instantaneous temporal evolution of structures.

Consider a given state in the $k$th snapshot. To find the future counterparts of the given state, we iterate through all states in the subsequent snapshot and determine which ones have any constituent points that are spatially-connected to the given state. In other words, we look for states with points that coincide with or neighbor any of the given state's points (using the 26 nearest neighbor criterion). Any such states are identified as future counterparts of the given state.

Let the states in each snapshot be denoted by an index $i \in \{ 1, 2, \dots, N_k \}$, where $N_k$ is the number of states in the snapshot. We construct an array that stores the indices of the corresponding future states in the $(k+1)$th snapshot, which we call the \it forward evolution \rm map $T^+_{k}(i)$ where $i = 1,2,\dots,N_k$. Thus, if the $i$th state in snapshot $k$ is associated with the $j$th state in snapshot $k+1$, then $T^+_{k}(i) = j$. If the $i$th state has no future counterparts, then we assign $T^+_{k}(i) = 0$, representing the null state, which corresponds to the destruction of the structure. If the $i$th state is associated to multiple future states (i.e., it divides), then we assign $T^+_{k}(i)$ multiple values containing all of these future states.

We perform a similar procedure to find the past counterparts of the given state. In this case, we iterate through all states in the preceding snapshot to determine which ones have points that are spatially-connected to the given state. Any such states are identified as past counterparts of the given state. Likewise, we construct a \it backward evolution \rm map, $T^-_{k}(i)$, which identifies state $i$ in the $k$th snapshot with its corresponding past counterparts in the $(k-1)$th snapshot. If the $i$th state has no past counterpart, then we assign $T^-_{k}(i) = 0$, which corresponds to the formation of the structure. If the $i$th state is associated to multiple past states (i.e., it results from a merger), then we assign $T^-_{k}(i)$ multiple values containing all of these past states.

The entire set of evolution maps $T^\pm_{k}$ where $k = 1, \dots, N_\text{snap}$ is constructed by iterating through the snapshots and applying the above procedure immediately after identifying the states in pairs of consecutive snapshots. After $T^\pm_{k}$ is constructed for all $N_k$ states in a given snapshot $k$, the array containing the constituent points of the states is deleted to free up memory before loading the next snapshot.

\begin{figure}
\includegraphics[width=12cm]{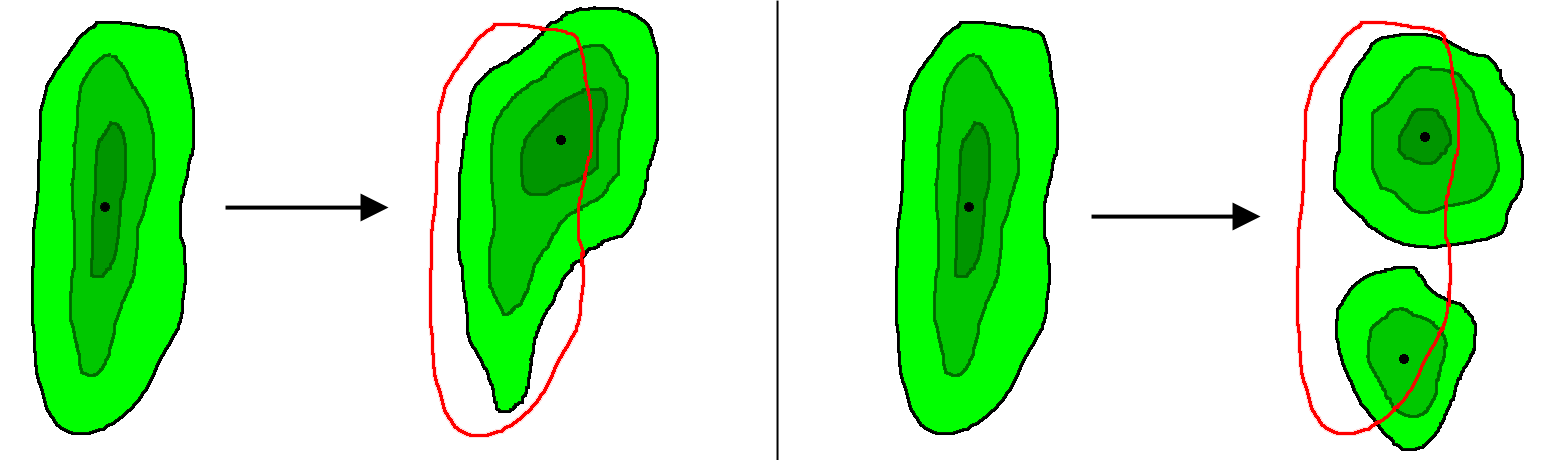}
   \centering
   \caption{\label{fig:evolve} Schematic of structure evolution, shown in 2D space for clarity. The procedure stores the constituent points of the present state, checks the future snapshot for any states with points that are spatially-connected to the present state's points, and then identifies these states as future counterparts of the present state. In the left panel, a structure evolves without interacting. In the right panel, a structure divides, having multiple future counterparts. This procedure is reversed in time to determine past counterparts, with a merger occuring for multiple past counterparts.}
 \end{figure}

\subsection{Identification of paths (tracking algorithm)}

In this subsection, we describe our algorithm for identifying paths across multiple snapshots, which is an important intermediate step before processes themselves can be identified. A path is abstractly defined as a sequence of states that is bijectively-connected under the evolution maps (see Sec.~\ref{class}). From the algorithms in the preceding subsections, we have a sequence of $N_\text{snap}$ snapshots (denoted by index $k$), each with a set of $N_k$ states, along with the evolution maps $T^\pm_{k}$ associating the states in each snapshot with their time-evolved counterparts in adjacent snapshots. To identify paths, we must find the bijective sequences in $\{ T^\pm_{k} \}$.

Consider the $i$th state in snapshot $k$. Bijectivity of the forward evolution map is satisfied if there exists a unique state $j$ in snapshot $k+1$ such that $T^+_{k}(i) = j$ and $T^-_{k+1}(j) = i$. Any two states satisfying this condition form a path segment. If this condition fails, however, then state $i$ is the endpoint of a path. This happens either when the structure is destroyed ($T^+_{k}(i) = 0$) or when it interacts (either $T^+_{k}(i)$ or $T^-_{k+1}(j)$ is multi-valued). Likewise, if there exists a unique state $l$ in snapshot $k-1$ such that $T^-_{k}(i) = l$ and $T^+_{k-1}(l) = i$, then these two states form a path segment; otherwise state $i$ is the beginning of a path due to either formation ($T^-_{k}(i) = 0$) or an interaction (either $T^-_{k}(i)$ or $T^+_{k-1}(l)$ is multi-valued). Using these conditions, we can track the state of any given structure along a path through a sequence of snapshots, until an interaction is encountered.

We iterate through the states in all $N_\text{snap}$ snapshots and use the above tracking procedure to identify the associated path and its constituent states, marking those states so that they are ignored in the remaining iterations. As a result, we obtain a set of $N_\text{path}$ paths. For each path, we construct an array which contains the indices of the constituent states, which is referenced to perform measurements on the path. We also construct an array containing the indices of the \it predecessors\rm, which are the other paths connecting to it from the beginning of the path. The predecessors are determined by operating with $T^-$ on the first state of the path to obtain all of the past states, and finding the paths that contain these past states. In a separate array, we store the indices of \it successors\rm, which are the other paths connecting to the end of the path, obtained by operating with $T^+$ on the final state of the path. The predecessors and successors of paths characterize the vertices between paths. Note that if the number of predecessors is zero, then the path is formed spontaneously. If the number of successors is zero, then the path is destroyed spontaneously.

\subsection{Identification of processes \label{sec:process_identification}}

We finally describe how to identify processes from the set of paths and their predecessors and successors. Recall that processes are described as sets of connected paths. Therefore, we first iterate through the set of paths. For each path, iterate through the predecessors and successors of the path, and then through the predecessors and successors of those, and so on, until no new paths can be obtained. The set of paths acquired in this way constitute a single process, and their indices are stored in an array corresponding to that process. The paths that have already been identified as belonging to a process are ignored in the remaining iterations of paths.

Some processes will contain states from the initial snapshot or in the final snapshot of the dataset, which we call \it initial \rm processes or \it final \rm processes, respectively. These processes must be treated as special cases, since our information about them is incomplete. The simplest treatments of these processes are either to ignore them or to treat them as normal processes undergoing formation or destruction in the initial or final snapshots. For most of our analysis, we will ignore initial and final processes. However, they are included in the probability distributions for better statistics at long durations. Due to the relatively long interval of time in our simulations, the initial and final processes are a very minor contribution to the statistics, unless low thresholds are used.

This concludes our discussion of the algorithms used to identify processes. We now have a sample of $N_\text{proc}$ processes, each including an array of constituent paths. The paths contain all of the information necessary to perform measurements on the processes. These measurements are described in the next section.

\section{Measurements} \label{sec4}

\subsection{Measurements for states}

In this subsection, we describe measurements made on states, i.e., spatial structures at fixed time, which will be used in the next subsections to construct similar, more general quantities for paths and processes.

The volume $V$ of the structure is immediately obtained by counting the number of constituent points of the state and multiplying by the lattice volume element. The Ohmic energy dissipation rate is given by
\begin{align}
{\cal E} = \int dV \eta j^2 \, ,
\end{align}
where integration is performed across all constituent points of the given state.

To characterize the geometry of each state, we measure the linear scales in three orthogonal directions. For length $L$, we take the maximum distance between any two points of the structure. For width $W$, we consider the plane orthogonal to the length and coinciding with the point of peak current density. We then take the maximum distance between any two constituent points in this plane to be the width. The direction for thickness $T$ is then fixed by the orthogonality condition. We take the thickness to be the distance across the structure in this direction through the point of peak current density. Since typical thicknesses tend to be comparable to the lattice spacing, we use a linear interpolation scheme to obtain finer measurements. Note that this method can cause the thickness to be over-estimated in some cases with complex morphologies, for example, in structures with S-shaped cross-sections, but these appear to be a relatively small population manifest as spurious high-value measurements. These definitions automatically satisfy $L \ge W \ge T$. 

We measure these scales in units of the system size in the direction perpendicular to the guide field. See \cite{zhdankin_etal2014} for more detailed information on this method applied to high-resolution simulations of MHD turbulence, as well as a discussion on alternative methods based on the Minkowski functionals.

States can also be characterized by some discrete properties. One example is the direction of the current flow, i.e., orientation. Another example is the presence/absence of topological features such as X-points or O-points [see \cite{zhdankin_etal2013} and \cite{servidio_etal2010}]. A final example is the Euler characteristic, i.e., genus. These characteristics will be relegated to a future paper.

\subsection{Measurements for paths}

In this subsection, we describe measurements for paths, which are conceptually simpler than those for processes due to the bijectivity condition. These measurements will be used and generalized in the next subsection to characterize processes.

The evolution of a path can be described by the time-series ${\cal{E} \rm}(t)$, $V(t)$, $L(t)$, $W(t)$, and $T(t)$ of instantaneous characteristics defined in the previous section for the constituent states. Consider a path given by a sequence of states at times $t_k$, where $k \in \{1,\dots,N_\text{s}\}$ and $N_\text{s}$ is the number of constituent states. Assuming a fixed cadence of snapshots, the states are separated by a fixed time interval $\Delta{t}$. The $k$th state has characteristics denoted by ${\cal E}_k$, $V_k$, $L_k$, $W_k$, and $T_k$.

One of the most basic properties of a path is its duration (or lifetime) $\tau$, defined as
\begin{align}
\tau = t_{N_\text{s}} - t_{1} = (N_s - 1) \Delta{t} \, .
\end{align}
The energy dissipation rate of states generalizes to the dissipated energy,
\begin{eqnarray}
E = \int dt {\cal{E}}(t) = \sum_{k = 1}^{N_s} {\cal E}_k \Delta{t} \, ,
\end{eqnarray}
where time integration is performed over the duration of the path. We also define the peak volume, peak energy dissipation rate, maximum length, maximum width, and maximum thickness as
\begin{eqnarray}
V_{\rm max} &=& {\rm max}(V(t)) = \max(\{ V_k \}) \nonumber \\
{\cal{E}}_{\rm max} &=& {\rm max}({\cal{E}}(t)) = \max(\{ {\cal E}_k \})  \nonumber \\
L_\text{max} &=& \max{(L(t))} = \max(\{ L_k \}) \nonumber\\
W_\text{max} &=& \max{(W(t))} = \max(\{ W_k \}) \nonumber\\
T_\text{max} &=& \max{(T(t))} = \max(\{ T_k \}) 
\end{eqnarray}
Note that since these quantities are local rather than time-integrated, they may be sensitive to chaotic fluctuations. As an alternative, we can consider time-averaged quantities, which however are less easily generalized for processes.

\subsection{Measurements for processes \label{sec:process_characteristic}}

We now describe how to generalize the quantities defined for paths in the previous section to processes, i.e., sets of interacting structures. We characterize each process by the number of constituent paths, $N_p$. Processes with a single path, $N_p = 1$, are isolated structures. Processes with three paths, $N_p = 3$, are division or merger processes. Processes consisting of more than three paths, $N_p > 3$, are higher-order processes, containing either more than one vertex or vertices joining more than three paths. Other related measures of the complexity of a process include the number of vertices $N_v$, the number of constituent states $N_s$, the number of internal paths $N_\text{int}$ (i.e. paths that begin and end in vertices), and the number of incoming paths $N_\text{in}$ or outgoing paths~$N_\text{out}$.

Consider a process with constituent paths enumerated by index $n = 1,\dots,N_p$, each extending from initial times $t_n$ to final times $t'_n$. Let $E_n$, $V_{{\rm max},n}$, ${\cal E}_{{\rm max},n}$, $L_{\text{max},n}$, $W_{\text{max},n}$, and $L_{\text{max},n}$ be the characteristics of the $n$th path as defined in the previous subsection. We define the process duration by
\begin{align}
\tau = \max (\{t'_n\}) - \min (\{t_n\}) \, ,
\end{align}
which is simply the time interval from the formation of the first existing constituent path to the destruction of the last existing constituent path. The total dissipated energy $E$ of a process is defined by integrating $\eta j^2$ across the enclosed 4D spacetime region. This is found in a straightforward manner from
\begin{eqnarray}
E = \sum_{n=1}^{N_p} E_{n} \, .
\end{eqnarray}
Likewise, we can define the peak volume $V_\text{max}$ and peak energy dissipation rate ${\cal E}_\text{max}$ as the maximum of the peaks corresponding to the constituent paths,
\begin{eqnarray}
V_{\rm max} = \max (\{ V_{{\rm max},n} \}) \nonumber \\
{\cal{E}}_{\rm max} =  \max (\{ {\cal E}_{{\rm max},n} \}) \, .
\end{eqnarray}
Note that an alternative definition for $V_\text{max}$ (${\cal E}_\text{max}$) can be based on the maximum of the volume (energy dissipation rate) summed for all states belonging to the process at any given time. These two definitions may differ for processes with a large number of paths, but will otherwise give similar results; we use the first definition only for simplicity. The most difficult quantities to generalize for a process are the characteristic spatial scales. There appears to be no universally satisfactory way to obtain \it average \rm characteristic scales for a general configuration with many paths. If we apply an average across all paths (or all states) constituting the process, then the result may be skewed toward unphysical short-lived paths (or states). A simpler alternative is to take the maximum scale corresponding to any path in the process,
\begin{eqnarray}
L_\text{max} = \max (\{L_{\text{max},n}\}) \nonumber \\
W_\text{max} = \max (\{W_{\text{max},n}\}) \nonumber \\
T_\text{max} = \max (\{T_{\text{max},n}\}) \, .
\end{eqnarray}
One alternative definition for the spatial scales, which gives results consistent with the definition that we apply, is to take the spatial scales from the largest state at the moment of peak energy dissipation rate.

\section{Results} \label{sec5}

\subsection{Simulation details}

Before discussing the results of our temporal analysis of MHD turbulence, we first describe the  numerical simulations. We consider simulations of strong, incompressible MHD turbulence driven at large scales. These simulations solve the reduced MHD equations, Eq. \ref{rmhd-elsasser}, using a fully dealiased 3D pseudo-spectral algorithm (see \citealt{perez_etal2012} for specific details on simulations). The ratio of magnetic guide field to rms fluctuations is set to $B_0/b_\text{rms} \approx 5$. The periodic box is elongated in $\hat{z}$ (the direction of the guide field) by a factor of $L_\parallel/L_\perp = 6$, where $L_\perp = 2\pi$ is the perpendicular size of the domain in simulation units. Turbulence is driven at the largest scales by colliding Alfv\'en modes, generated from statistically independent random forces $\boldsymbol{f}^\pm$ in Fourier space at low wave-numbers $2\pi/L_{\perp} \leq k_{x,y} \leq 2 (2\pi/L_{\perp})$, $k_z = 2\pi/L_\|$. The Fourier coefficients of $\boldsymbol{f}^\pm$ are Gaussian random numbers with amplitudes chosen so that $b_\text{rms}\sim v_\text{rms}\sim 1$. The forcing is solenoidal in the perpendicular plane and has no component along $\boldsymbol{B}_0$. The random values of the different Fourier components of the forces are refreshed independently on average about $10$ times per eddy turnover time. The Reynolds number is defined by $Re = v_{\rm rms} (L_\perp/2\pi) / \nu$; we set $\nu = \eta$ so that the magnetic Reynolds number, $Rm = v_{\rm rms} (L_\perp/2\pi) / \eta$, is equal to $Re$. Durations (and other timescales) are measured in terms of large-scale eddy turnover times of the turbulence, given by $\tau_\text{eddy} = L_\perp/(2 \pi v_\text{rms}) \approx 1$. The analysis is performed on time intervals of durations $\tau_\text{tot}$, all of which begin after the simulations reach statistical steady state.

\begin{table*}[h!b!p!]
\caption{List of numerical simulations \newline} \label{sims}
\begin{tabular}{|c|c|c|c|c|} 
	\hline
\hspace{0.5 mm} Case \hspace{0.5 mm}  & \hspace{2 mm} Resolution \hspace{2 mm}   & \hspace{4 mm}$Re$\hspace{4 mm}  &   \hspace{2 mm} $\Delta{t}$ \hspace{2 mm} &   \hspace{2 mm} $\tau_\text{tot}$ \hspace{2 mm} \\
	\hline
$1$ & $256^3$ & 800 & $1/64$ & 10.0  \\
$2$ & $512^3$ & 800 & $1/32$ & 12.2  \\
$3$ & $512^3$ & 1250 & $1/64$ & 15.6 \\
$4$ & $512^3$ & 1800 & $1/32$ & 12.2 \\
	\hline
\end{tabular}
\centering
\label{table-sims}
\end{table*}

We consider four simulations with parameters shown in Table~\ref{table-sims}. Case 1 is a lower-resolution ($256^3$) run used to establish convergence of the results with resolution. Three independent runs with resolution $512^3$ are chosen to study scalings with $Re$, although the relatively limited range ($Re = 800-1800$) inhibits precise measurements of the scalings. Of these runs, Case 3 with $Re = 1250$ is the most robust data set, having the highest cadence, being nominally well-resolved dynamically, and having the longest time interval ($\tau_\text{tot} = 15.6$).

We analyze snapshots dumped at a cadence $(\Delta t)^{-1}$, with $\Delta{t}$ being larger than the internal time step in the simulation. For reference, we now establish the naive estimate for the minimum cadence required to properly track structures between two adjacent snapshots. This is estimated by requiring that the distance advected by the flow during $\Delta{t}$ is less than the typical current sheet thickness. Estimating the former as $v_\text{rms} \Delta{t}$ and the latter as $b_\text{rms}/j_\text{thr}$, we require $\Delta{t} < b_\text{rms} / (v_\text{rms} j_\text{thr}) \approx 1/j_\text{thr}$. For the four cases in Table~\ref{table-sims}, we have $j_\text{rms} \in \{ 12.1, 11.8, 14.6, 17.4\}$, which gives the condition $\Delta{t} < \{ 1/12.1, 1/11.8, 1/14.6, 1/17.4 \} j_\text{rms}/j_\text{thr}$. The cadences given in Table~\ref{table-sims} fall short of satisfying this condition for thresholds more than three or four times larger than the rms fluctuations. However, we note that this condition is somewhat alleviated in practice because associations are made for states that do not fully overlap (since displacements of one lattice spacing satisfy the connectivity condition) and the condition is applied to all of the points in the (generally large) structure. Most of the results in our analysis show robust convergence with cadence or are only weakly sensitive.

\subsection{Global results}

\begin{figure}
\includegraphics[width=8cm]{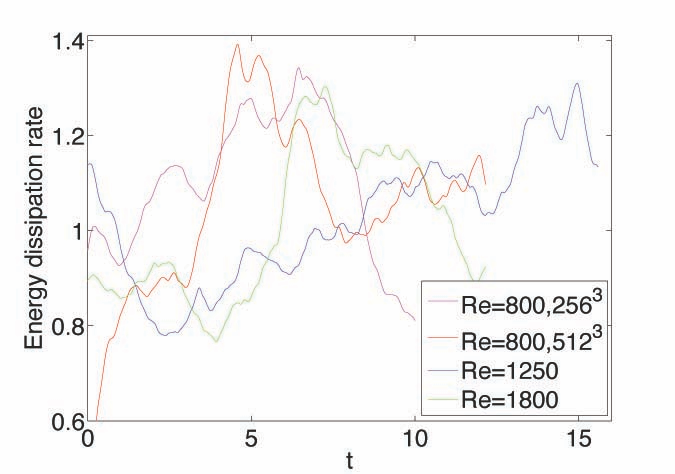}
\includegraphics[width=8cm]{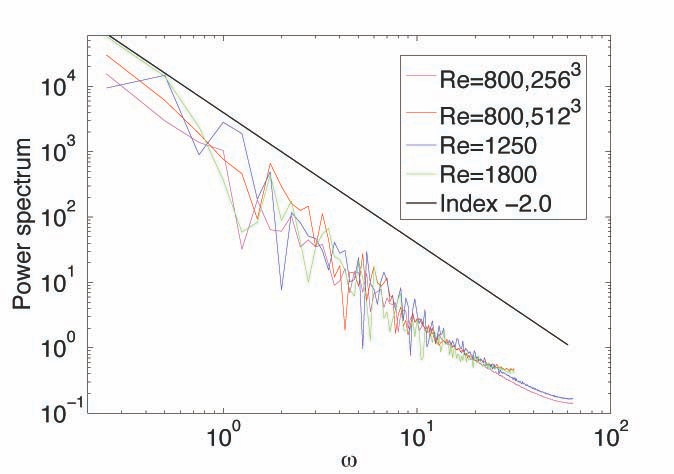}
   \centering
   \caption{\label{fig:lightcurve} Left panel: the total energy dissipation rate in the system for the analyzed interval of time in the four simulations. Right panel: the power spectrum of this time series, showing a power law index near -2. The colors correspond to cases 1 (magneta), 2 (red), 3 (blue), and 4 (green) from Table~\ref{table-sims}.}
 \end{figure}

We first describe the global features of energy dissipation in the simulations. In the first panel of Fig.~\ref{fig:lightcurve}, we show the total ohmic energy dissipation rate in the system, ${\cal E}_\text{tot}(t)$, during the given time intervals for all of the simulations in Table~\ref{table-sims}. If we associate the dissipated energy with prompt optically-thin emission, i.e., if we assume that all dissipated energy is converted immediately into radiation in an optically-thin environment, then this represents a light curve for the system. The mean of ${\cal E}_\text{tot}(t)$ is very close to $1.0$, in agreement with the energy input from the large-scale forcing. The rms fluctuation about this mean is approximately $0.15$ for all cases. In the second panel of Fig.~\ref{fig:lightcurve}, we show the power spectra of ${\cal E}_\text{tot}(t)$, which exhibits a power law with index close to $-2.0$ for all cases.

\begin{figure}
\includegraphics[width=8cm]{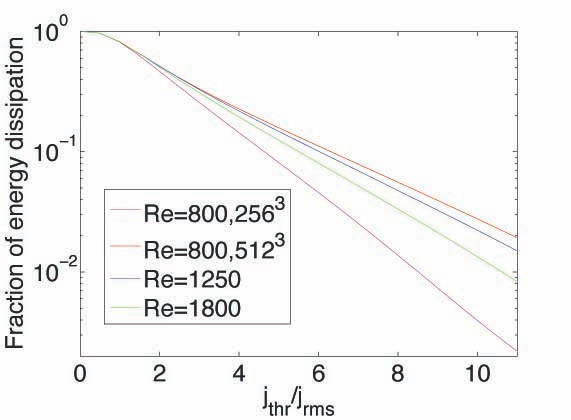}
\includegraphics[width=8cm]{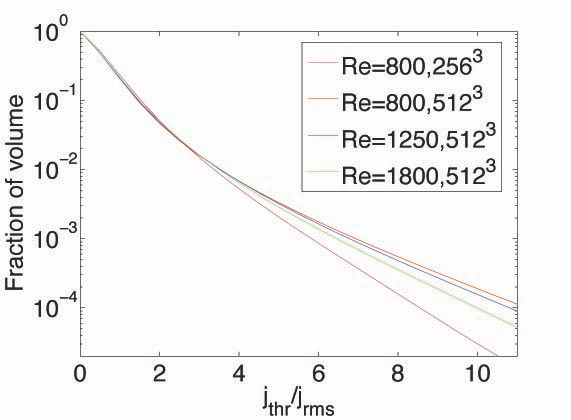}
   \centering
   \caption{\label{fig:fractions} The fractions of total energy dissipation (left) and volume (right) in structures with $|j| > j_\text{thr}$. The curves correspond to cases 1 (magenta), 2 (red), 3 (blue), and 4 (green) from Table~\ref{table-sims}. Case 2 is consistent with the result from fully-resolved simulations obtained in \cite{zhdankin_etal2014}.}
 \end{figure}

We now consider the energetics of structures with varying thresholds (normalized by the rms current density for the entire time interval, noting that $j_\text{rms} = \sqrt{{\cal{E}}_\text{tot} / \eta V_\text{tot}} \propto Rm^{1/2}$). The fraction of total energy dissipation occurring in structures with $|j| > j_\text{thr}$, given by $\int_{j_\text{thr}}^\infty dj P(j) \eta j^2 / (\int_0^\infty dj P(j) \eta j^2)$, where $P(j)$ is the probably distribution of $j$, is shown in Fig.~\ref{fig:fractions}. We find that this function has a wide tail characteristic of intermittency, declining exponentially at large $j_\text{thr}$. We also show the fraction of total volume occupied by the same structures, given by $\int_{j_\text{thr}}^\infty dj P(j)$, in the second panel of Fig.~\ref{fig:fractions}. The fraction of energy dissipation always greatly exceeds the occupied volume, with, for example, about $30\%$ of the energy dissipation occurring in $1\%$ of the volume at $j_\text{thr}/j_\text{rms} \approx 3$. Our previous study has shown that the fractions of energy dissipation and occupied volume both may be universal for sufficiently well-resolved, high-$Re$ simulations \citep{zhdankin_etal2014}. Cases 2 and 3 closely match these previous results. Since significant deviations occur at large $j_\text{thr}/j_\text{rms}$ for the other cases, they may not be completely well-resolved.

\begin{figure}
\includegraphics[width=16cm]{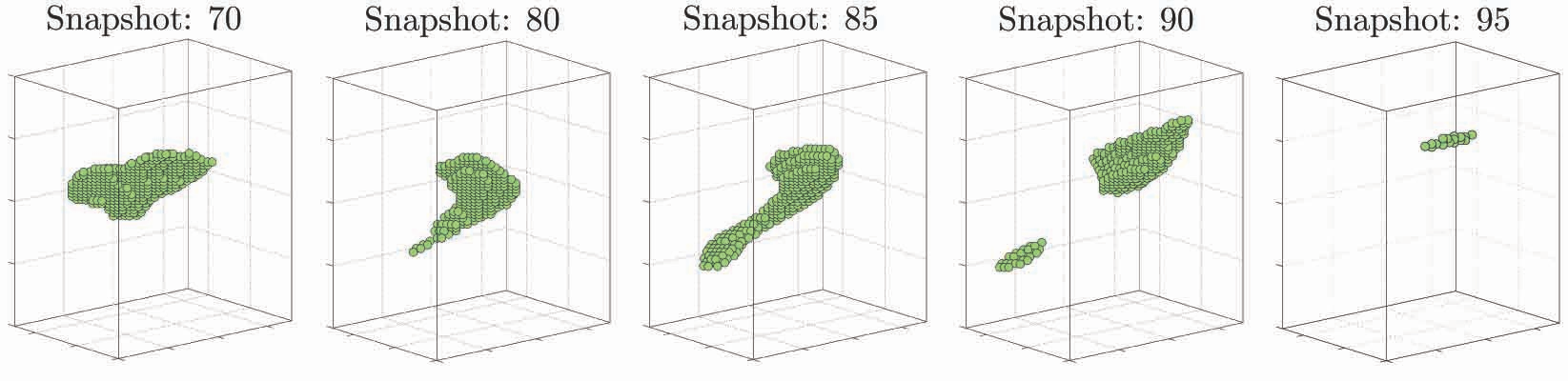}
\includegraphics[width=12cm]{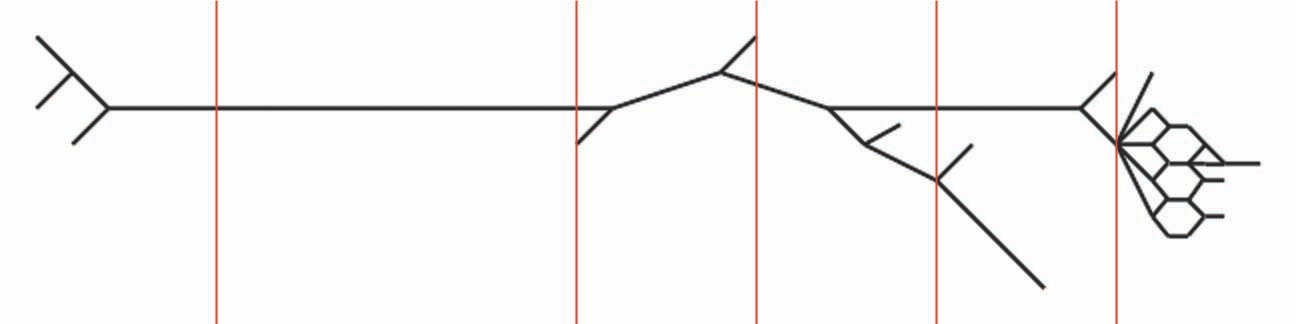}
   \centering
   \caption{\label{fig:tree} An example of a typical process with duration $\tau \approx 0.5$, shown in green on the simulation lattice. A schematic diagram of the process is also shown, with the snapshot times marked by a red line.}
 \end{figure}

Before proceeding to the quantitative analysis of processes in the simulations, we show an example of a relatively simple process in Fig.~\ref{fig:tree}. This process has a duration $\tau \approx 0.5$ with $31$ distinct paths. Snapshots of a few representative states (in green) are shown on a subdomain of the simulation lattice (with size $0.10 \times 0.14 \times 0.90$). These images do not account for the elongation of the lattice along the (vertical) guide field, which would emphasize the ribbon-like character of the structures. The broadest part of the structure is shown from the given perspective. In addition to these snapshots, we also show the schematic diagram of the process (as established in Sec.~\ref{class}), with the snapshot times marked in red. It is evident that this process is highlighted by a major division which occurs after the structure is stretched. A disproportionally large number of paths in this case are produced at the final stages of the process, as it decays toward the threshold.

\subsection{Aggregate quantities}

We now consider some aggregate quantities for the sample of processes in our simulations.  Note that occurrence rates are generally not a very robust statistic, since they can be strongly affected by structures near the threshold, which in turn are strongly affected by resolution and cadence. A more robust analysis would filter out the small-scale and under-resolved structures to circumvent these numerical issues. We use no filtering in the present analysis, both for simplicity and because we are only interested in broad trends.

\begin{table*}[h!b!p!]
\caption{Aggregate quantities in four simulations ($j_\text{thr}/j_\text{rms} \approx 6.8$, $\Delta{t} = 1/32$) \newline}
\begin{tabular}{|c|c|c|c|c|} 
	\hline
\hspace{0.5 mm} Quantity \hspace{0.5 mm}  & \hspace{2 mm}$256^3, Re = 800$\hspace{2 mm}    &   \hspace{2 mm}$512^3, Re = 800$\hspace{2 mm} &\hspace{2 mm}$512^3, Re = 1250$  & \hspace{2 mm}$512^3, Re = 1800$\hspace{2 mm}  \hspace{2 mm}\\
	\hline
$\langle N_\text{state} \rangle$ & 194 & 288 & 657 & 1328  \\
$N_\text{proc}$ & 914 & 1271 & 4272 & 11608 \\
$N_\text{proc}/N_\text{path}$ & 0.218 & 0.240 & 0.278 & 0.339 \\
$N_\text{int}/N_\text{isol}$ & 0.278 & 0.329  & 0.187 & 0.115 \\
$N_\text{mer}/N_\text{div}$ & 0.839 & 0.776  & 0.800 & 0.823 \\ 
$N_\text{form}/N_\text{des}$ & 0.871 & 0.853 & 0.886 & 0.911 \\
$N_\text{n-vert}/N_\text{3-vert}$ & 0.341 & 0.353 & 0.485 & 0.523  \\
	\hline
\end{tabular}
\centering
\label{table-cases}
\end{table*}

Some general results are shown in Table~\ref{table-cases} for all of the simulations with fixed threshold $j_\text{thr}/j_\text{rms} \approx 6.8$ and cadence $\Delta{t}^{-1} = 32$. Here, $N_\text{proc}$ is the number of processes, $N_\text{path}$ is the number of paths, $N_\text{int}$ is the number of processes with interactions (i.e., processes consisting of more than one path), $N_\text{isol}$ is the number of isolated structures (i.e., processes consisting of one path), $N_\text{mer}$ is the number of merger events (where a vertex with $n$ ingoing paths counts as $n-1$ mergers), $N_\text{div}$ is the number of division events (where a vertex with $n$ outgoing paths counts as $n-1$ divisions), $N_\text{form}$ is the number of formation events (i.e., the number of paths with no predecessors), $N_\text{des}$ is the number of destruction events (i.e., the number of paths with no successors), $N_\text{3-vert}$ is the number of three-point vertices, and $N_\text{n-vert}$ is the number of $n$-point vertices with $n > 3$. All of these preceding quantities are normalized to the number per eddy turnover time. We also show the mean number of states per snapshot, $\langle N_\text{state} \rangle$. Table~\ref{table-cases} mainly shows the ratios of these various quantites; see the much more detailed Table~\ref{table-cases2} in Appendix~\ref{appendix:tables} for actual occurrence rates.

We now make several remarks about Table~\ref{table-cases}. First, $\langle N_\text{state} \rangle$ and $N_\text{proc}$ in Case 1 are smaller than the corresponding values for the higher-resolution Case 2, with a discrepancy of about $30\%$. In fact, there is a similar discrepancy in the number of occurrences for all measured quantities (see Table~\ref{table-cases2} in Appendix). This can be attributed to the fact that Case 1 is somewhat under-resolved and therefore misses some of the small-scale dynamics present in Case 2. Second, $\langle N_\text{state} \rangle$ and $N_\text{proc}$ both strongly increase with $Re$, obeying estimated scalings of $\langle N_\text{state} \rangle \sim Re^{1.9}$ and $N_\text{proc} \sim Re^{2.7}$. Third, three-point vertices are more common than higher-order vertices, although the higher-order vertices still occur in significant numbers. Higher-order vertices appear to be an unavoidable consequence of time discretization. Indeed, the ratio $N_\text{n-vert}/N_\text{3-vert}$ increases with $Re$, implying that interactions occur over smaller timescales for higher $Re$. Fourth, although there are fewer processes with interactions than processes with no interactions, they contain the majority of the paths (i.e., $N_\text{proc}/N_\text{path} < 1/2$). Fifth, divisions are somewhat more common than mergers, implying a time asymmetry in the interactions. Equivalently, there are more destructions than formations. This asymmetry is reasonable since the time-reversal symmetry of the ideal MHD equations is broken by resistive and viscous dissipation. The preference for divisions over mergers may be a manifestation of the direct cascade of energy from large scales to small scales.

We next consider the effect of cadence, $\Delta{t}^{-1}$, on the above quantities. We show the various quantities from Case 3 for $1/64 \le \Delta{t} \le 1/4$ in Table~\ref{table-cadence}. Most quantities monotonically increase when the cadence is increased, or equivalently, $\Delta{t}$ is decreased (see Table~\ref{table-cadence2} and Table~\ref{table-cadence3} in Appendix~\ref{appendix:tables}). However, $N_\text{proc}$ first increases then decreases with cadence, showing a local maximum near $\Delta{t} \approx 1/16$. This unintuitive result may be explained as follows. Although the number of paths always increases with cadence, the connectivity of paths changes. For low cadence, paths tend to be isolated structures. This may be a sign that the cadence is insufficient to properly track structures - in particular, in the limit of very large $\Delta{t}$, snapshots become completely uncorrelated, so most processes appear as single states, therefore $N_\text{proc} \sim (\tau_\text{tot}/\Delta{t}) \langle N_\text{state} \rangle$. For intermediate cadence (i.e., once structures are properly tracked), the paths interact more often when cadence increases, eventually decreasing $N_\text{proc}$ due to the combining of isolated structures. The transition between these two regimes can be expected when $N_\text{proc}/N_\text{path} \approx 1/2$, which is close to the local peak in $N_\text{proc}$ near $\Delta{t} \approx 1/16$ for this case. For very high cadence, $N_\text{proc}$ either saturates or increases once again. The latter scenario may occur if additional isolated structures appear at the shortest time-scales, or if the processes are fractal. This trend is observed for Case 1 (see Table~\ref{table-cadence2} in Appendix~\ref{appendix:tables}).

\begin{table*}[h!b!p!]
\caption{Variation of select aggregate quantities with cadence (Case 3: $512^3, Re=1250, j_\text{thr}/j_\text{rms} \approx 6.8$) \newline}
\begin{tabular}{|c|c|c|c|c|c|} 
	\hline
\hspace{0.5 mm} Quantity \hspace{0.5 mm}  & \hspace{0.2 mm} $\Delta{t} = 1/64$ \hspace{0.2 mm}  & \hspace{0.2 mm}  $\Delta{t} = 1/32$ \hspace{0.2 mm} & \hspace{0.2 mm} $\Delta{t} = 1/16$ \hspace{0.2 mm} & \hspace{0.2 mm} $\Delta{t} = 1/8$ \hspace{0.2 mm}   & \hspace{0.2 mm} $\Delta{t} = 1/4$ \hspace{0.2 mm}  \\
	\hline
$N_\text{proc}$ & 3311 & 4272 & 4908 & 3704 & 2197 \\
$N_\text{proc}/N_\text{path}$ & 0.120 & 0.278 & 0.562 & 0.780 & 0.895 \\
$N_\text{int}/N_\text{isol}$ & 0.429 & 0.176 & 0.067 & 0.030 & 0.016 \\
$N_\text{mer}/N_\text{div}$ & 0.845 & 0.799 & 0.797 & 0.771 & 0.778 \\
$N_\text{form}/N_\text{des}$ & 0.831 & 0.886 & 0.953 & 0.978 & 0.992 \\ 
$N_\text{n-vert}/N_\text{3-vert}$ & 0.367 & 0.485 & 0.607 & 0.697 & 0.772 \\ 
	\hline
\end{tabular}
\centering
\label{table-cadence}
\end{table*}

Finally, we consider the effect of the threshold, $j_\text{thr}$, on the above quantities. We show the various quantities from Case 3 for $5.5 \le j_\text{thr}/j_\text{rms} \le 9.6$ in Table~\ref{table-threshold}. For these relatively high thresholds, the occurrence rates for all quantities increase as $j_\text{thr}$ decreases, due to the larger sample size of paths (see Table~\ref{table-threshold2} in Appendix~\ref{appendix:tables}). In particular, we estimate that $\langle N_\text{state} \rangle \sim j_\text{thr}^{-3.5}$ and $N_\text{path} \sim j_\text{thr}^{-3.1}$, although these must deviate from a power law as $j_\text{thr} \to j_\text{rms}$. In contrast, the ratios of the occurrence rates change relatively little. The asymmetry of interactions decreases and the relative proportion of isolated structures increases as $j_\text{thr}$ decreases, likely due to a larger sample of small structures near the threshold.

The onset of percolation occurs near $j_\text{thr}/j_\text{rms} \approx 6.8$ for structures through space (in the periodic domain along the guide field, when $\max{\{L\}} \to L_\parallel/(2\pi)$), and near $j_\text{thr}/j_\text{rms} \approx 5.5$ for processes through the given time interval (when $\max{\{\tau\}} \to \tau_\text{tot}$). We call this latter quantity the \it percolation threshold\rm, which, in the limit $\tau_\text{tot} \to \infty$, is a fundamental characteristic current density of the system. For $j_\text{thr}$ below the percolation threshold, the initial processes and final processes (defined in Sec.~\ref{sec:process_identification}) contain a large fraction of the dissipated energy, as can be seen in Table~\ref{table-threshold} from $E_\text{interior}/E_\text{all}$, which is the ratio of energy dissipated by interior processes (i.e., processes that contain no states from the initial or final snapshots), $E_\text{interior}$, to energy dissipated by all processes (including initial and final processes), $E_\text{all}$. This ratio is large (i.e., $E_\text{interior}/E_\text{all} > 0.7$) until the percolation threshold is approached near $j_\text{thr}/j_\text{rms} \approx 5.5$, where the ratio quickly becomes small ($E_\text{interior}/E_\text{all} \sim 0.29$). The percolation threshold sets a practical limit on the smallest threshold for a reliable temporal analysis, since percolation otherwise interferes with the statistics of structures at the largest scales.

\begin{table*}[h!b!p!]
\caption{Variation of select aggregate quantities with changing threshold (Case 3: $512^3, Re = 1250$, $\Delta{t} = 1/64$) \newline}
\begin{tabular}{|c|c|c|c|c|} 
	\hline
\hspace{0.5 mm} Quantity \hspace{0.5 mm}  & \hspace{0.2 mm} $j_\text{thr}/j_\text{rms} \approx 9.6$ \hspace{0.2 mm}  & \hspace{0.2 mm}  $j_\text{thr}/j_\text{rms} \approx 8.2$ \hspace{0.2 mm} & \hspace{0.2 mm} $j_\text{thr}/j_\text{rms} \approx 6.8$ \hspace{0.2 mm} & \hspace{0.2 mm} $j_\text{thr}/j_\text{rms} \approx 5.5$ \hspace{0.2 mm}  \\
	\hline
$\langle N_\text{state} \rangle$ & 190 & 343 & 657 & 1287  \\ 
$N_\text{proc}$ & 1105 & 1777 & 3311 & 6314  \\
$N_\text{proc}/N_\text{path}$ & 0.139 & 0.125 & 0.120 & 0.116  \\
$N_\text{int}/N_\text{isol}$ & 0.491 & 0.481 & 0.429 & 0.383  \\
$N_\text{mer}/N_\text{div}$ & 0.825 & 0.832 & 0.845 & 0.863  \\
$N_\text{form}/N_\text{des}$ & 0.820 & 0.820 & 0.831 & 0.869 \\ 
$N_\text{n-vert}/N_\text{3-vert}$ & 0.313 & 0.331 & 0.367 & 0.400 \\  
$E_\text{interior}/E_\text{all}$ & 0.88 & 0.86 & 0.70 & 0.29  \\ 
$\max{\{L\}}$ & 2.6 & 3.3 & 6.0 & 6.0  \\ 
$\max{\{W\}}$ & 0.32 & 0.33 & 0.44 & 0.55  \\ 
$\max{\{\tau\}}$ & 3.3 & 3.9 & 8.5 & 13.8  \\ 
	\hline
\end{tabular}
\centering
\label{table-threshold}
\end{table*}

\subsection{Probability distributions and scaling relations} \label{sec:prob}

We now describe the probability distributions for the process characteristics, defined in Sec.~\ref{sec:process_characteristic}. For clarity, we focus on Cases 2-4, which have resolution $512^3$ and varying $Re$. We choose a relatively high threshold of $j_\text{thr}/j_\text{rms} \approx 6.8$, which is well above the percolation threshold. We retain initial and final processes for better statistics. The distributions are converged with respect to cadence and resolution (based on a comparison between Case 1 and Case 2 in Table~\ref{table-sims}), although it is possible that they have not fully converged with $Re$.

\begin{figure}
\includegraphics[width=5.4cm]{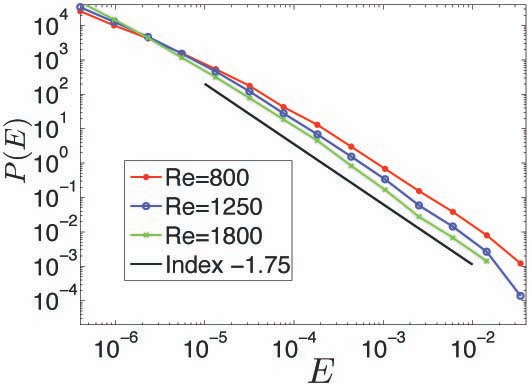}
\includegraphics[width=5.4cm]{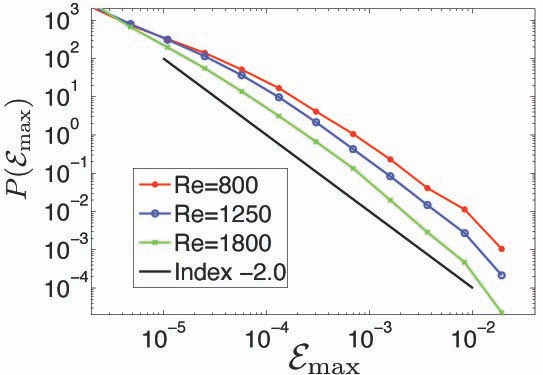}
\includegraphics[width=5.4cm]{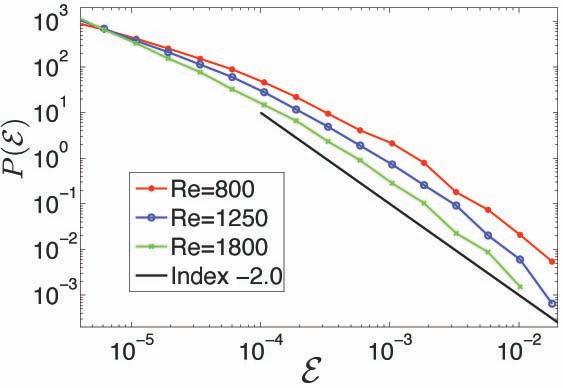}
   \centering
   \caption{\label{fig:dist_energy} The distributions for dissipated energy $E$, peak energy dissipation rate ${\cal E}_\text{max}$, and energy dissipation rate (of states) ${\cal E}$. These have power laws with index close to $-1.75$ for $P(E)$ and $-2.0$ for $P({\cal E}_\text{max})$ and $P({\cal E})$. The curves correspond to $Re = 800$ (red), $Re = 1250$ (blue), and $Re = 1800$ (green).}
 \end{figure}

We begin with the distribution for dissipated energy, $P(E)$, shown in the first panel of Fig.~\ref{fig:dist_energy}. We find that $P(E)$ has a power law tail with index near $-1.75\pm0.1$. The power law region extends across approximately three orders of magnitude in $E$, from $E \approx 10^{-5}$ up to about $E \approx 10^{-2}$. For smaller $E$, the distribution is shallower and appears to be non-universal, likely due to a combination of dissipation-range effects and threshold effects. With increasing $Re$, the power law extends to smaller $E$, consistent with a longer inertial range. If we instead consider the distribution for dissipated energy in isolated structures or in paths alone, then there is no clear power law.

The distribution for the peak energy dissipation rate, $P({\cal E}_\text{max})$, is shown in the second panel of Fig.~\ref{fig:dist_energy}. We find that $P({\cal E}_\text{max})$ has a power law with index close to~$-2.0 \pm 0.1$ in the range ${\cal E}_\text{max} \approx 10^{-4}$ to ${\cal E}_\text{max} \approx 10^{-2}$. Incidentally, this index is also observed in the distribution for energy dissipation rates, $P({\cal E})$, obtained from the population of spatial structures, i.e., states, shown in the third panel of Fig.~\ref{fig:dist_energy}. The power law index of $-2.0$ for $P({\cal E})$ is in agreement with our previous analysis using much higher $Re$ \citep{zhdankin_etal2014}. It can be shown that, if one assumes that all processes are single paths that evolve with identical, rescaled functional forms, then the indices for $P({\cal E}_\text{max})$ and $P({\cal E})$ will in fact be equal. The assumption of identical evolution for processes of all durations is consistent with our results in Subsection~\ref{sec:evolve}. See Appendix~\ref{appendix:constraint} for an analytic derivation of this relation.

\begin{figure}
\includegraphics[width=5.4cm]{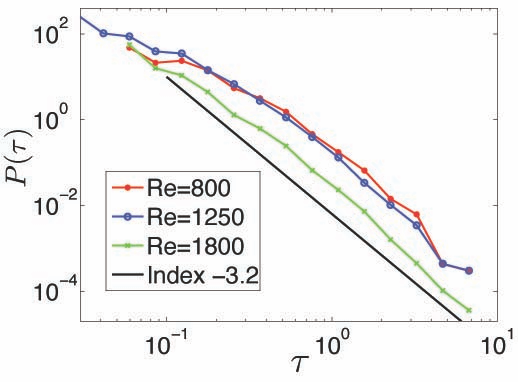}
\includegraphics[width=5.4cm]{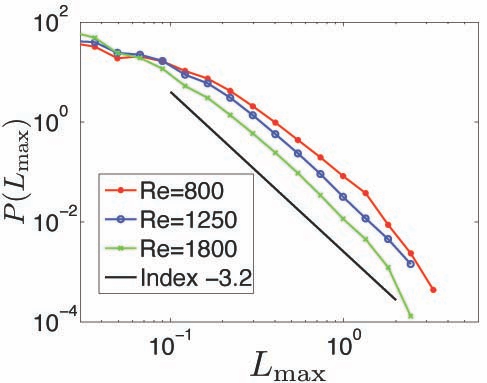}
\includegraphics[width=5.4cm]{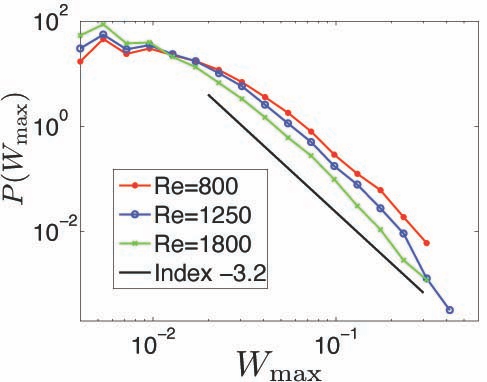}
   \centering
   \caption{\label{fig:dist_scales} The distributions for duration $\tau$, maximum length $L_\text{max}$, and maximum width $W_\text{max}$, all showing power laws with index near $-3.2$. The curves correspond to $Re = 800$ (red), $Re = 1250$ (blue), and $Re = 1800$ (green).}
 \end{figure}

The distribution for process duration $\tau$ is shown in the first panel of Fig.~\ref{fig:dist_scales}. The durations extend to well above an eddy turnover time, sometimes comparable to $\tau_\text{tot}$. The distribution from $\tau \approx 0.2$ to $\tau \approx 8$ can be fit by a power law with index near $-3.2 \pm 0.2$. Likewise, the distributions for maximum length $L_\text{max}$ and maximum width $W_\text{max}$ have power laws with indices near $-3.2$, also shown in Fig.~\ref{fig:dist_scales}. The distributions for all of these geometric quantities are related due to the strong correlations, described later in this subsection.

\begin{figure}
\includegraphics[width=6cm]{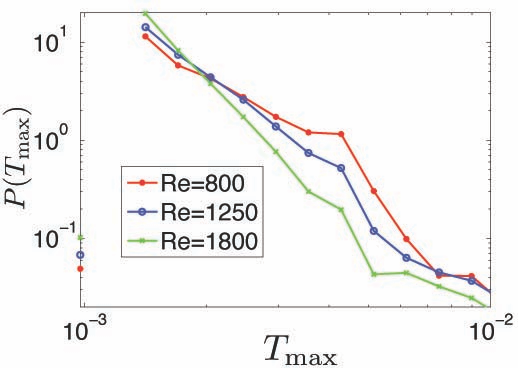}
\includegraphics[width=6cm]{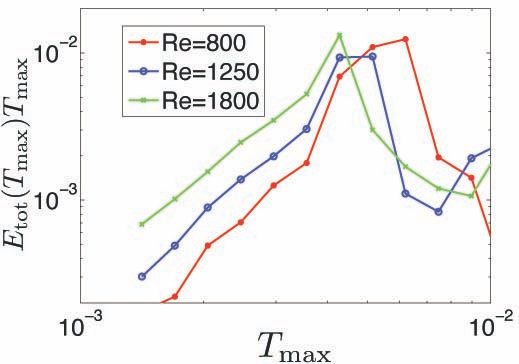}
   \centering
   \caption{\label{fig:dist_thick} Left panel: the distribution for maximum thickness $T_\text{max}$, which peaks at the lattice scale, $h \approx 0.02$. Right panel: the same distribution weighted by dissipated energy and compensated by $T_\text{max}$, showing that energy dissipation is dominated by processes with thicknesses a few times larger than the lattice scale. The curves correspond to $Re = 800$ (red), $Re = 1250$ (blue), and $Re = 1800$ (green).}
 \end{figure}

We now remark on the distribution for maximum thickness. As shown in the first panel of Fig.~\ref{fig:dist_thick}, $P(T_\text{max})$ peaks at the lattice scale, $h = 1/512 \approx 0.002$ (in units of system size, $L_\perp$). This is mainly due to the large population of under-resolved structures near the threshold. It is therefore more transparent to consider the distribution weighted by energy dissipation, i.e., to consider the total dissipated energy of processes at given $T_\text{max}$, which we denote $E_\text{tot}(T_\text{max})$. When $E_\text{tot}(T_\text{max})$ is compensated by $T_\text{max}$, the maximum corresponds to the thickness scale at which most of the energy dissipation occurs. As shown in Fig~\ref{fig:dist_thick}, $E_\text{tot}(T_\text{max})T_\text{max}$ is strongly peaked at $T_\text{max}$ a few times larger than the grid scale by (peaking at 2 to 3 times $h$, depending on $Re$).

The extremely small values and variation of thicknesses makes it challenging to accurately infer the corresponding scalings in the given simulations. Higher resolution simulations are needed in order to make proper measurements. However, it is reasonable to expect that the bulk features of the structures and processes are insensitive to the thickness. Indeed, the robust scalings of the other characteristics are strong evidence in favor of this. This is tied to the fact that the other characteristics are essentially inertial-range quantities, which should be insensitive to the dissipation dynamics (assuming locality of the energy cascade). Further evidence for this comes from higher resolution simulations with similar $Re$, which show that the thickness of states is concentrated at a similar value as found here \citep{zhdankin_etal2014}.

\begin{figure}
\includegraphics[width=7cm]{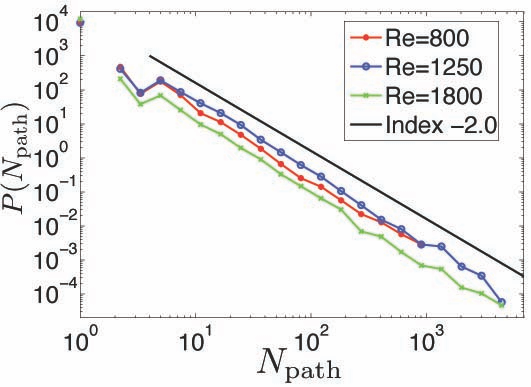}
   \centering
   \caption{\label{fig:dist_paths} The probability distribution for the number of paths per process, which is well fit by a power law with index $-2.0$. The most complex processes contain thousands of paths. The curves correspond to $Re = 800$ (red), $Re = 1250$ (blue), and $Re = 1800$ (green).}
 \end{figure}

Finally, we consider the distribution for the number of paths per process, $N_\text{path}$, which is shown in Fig.~\ref{fig:dist_paths}. We find that $P(N_\text{path})$ shows a robust power law with index near $-2.0 \pm 0.2$. The most complex processes have $\sim 10^3$ paths. It is remarkable that $P(N_\text{path})$  shows such a robust power law across nearly the entire range of values (roughly $1 < N_\text{path} < 4\times10^{3}$), since, a priori, it is not clear that the number of paths is a robust physical quantity.

\begin{figure}
\includegraphics[width=6cm]{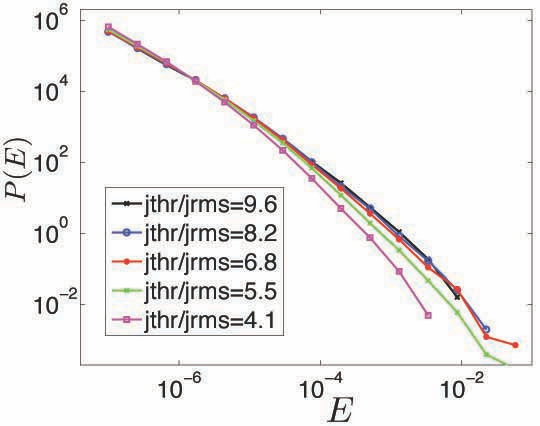}
\includegraphics[width=6cm]{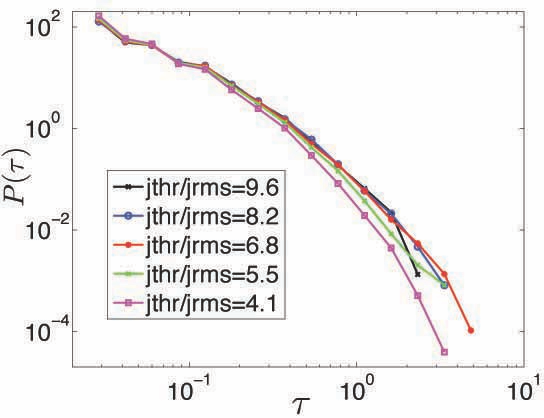}
   \centering
 \caption{\label{fig:dist_thresholds} The probability distributions $P({\cal E})$ (left) and $P(\tau)$ (right) at various $j_\text{thr}/j_\text{rms}$. The distributions are affected by percolation through the time interval for thresholds below $j_\text{thr}/j_\text{rms} = 5.5$ (green), visible in the curve for $j_\text{thr}/j_\text{rms} = 4.1$ (magenta).}
 \end{figure}

All of the distributions described above are insensitive to the threshold for large enough thresholds. As an example, we show $P({\cal E})$ and $P(\tau)$ for $4.1 \le j_\text{thr}/j_\text{rms} \le 9.6$ in Fig.~\ref{fig:dist_thresholds}. The distributions are similar in all cases with thresholds above the percolation threshold $j_\text{thr}/j_\text{rms} \approx 5.5$. Deviations in the tails of both distributions are discernable when $j_\text{thr}/j_\text{rms} = 5.5$ and are more evident when $j_\text{thr}/j_\text{rms} = 4.1$, well below the percolation threshold. The percolation of processes steepens the tails of the distributions, consistent with undercounting the large-scale processes.

\begin{figure}
\includegraphics[width=7cm]{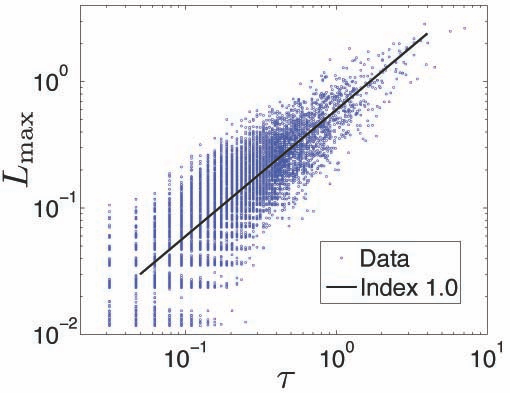}
\includegraphics[width=7cm]{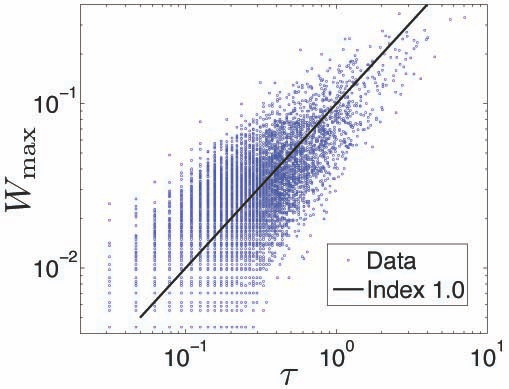}
\includegraphics[width=7cm]{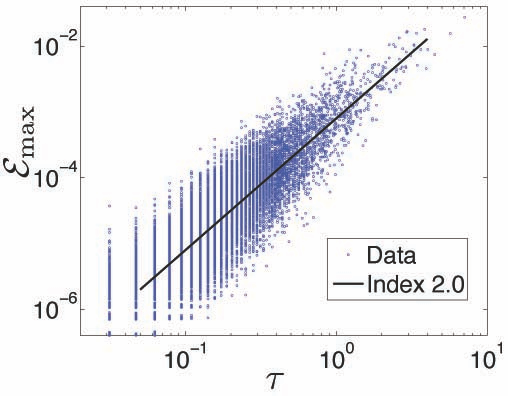}
\includegraphics[width=7cm]{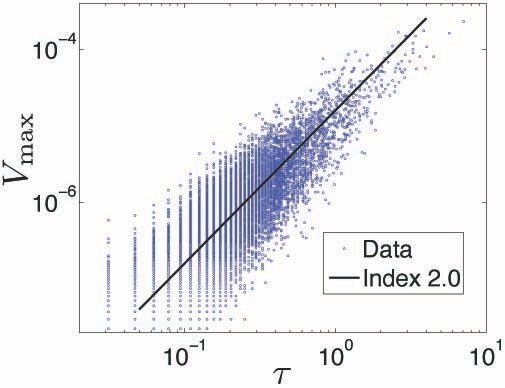}
\includegraphics[width=7cm]{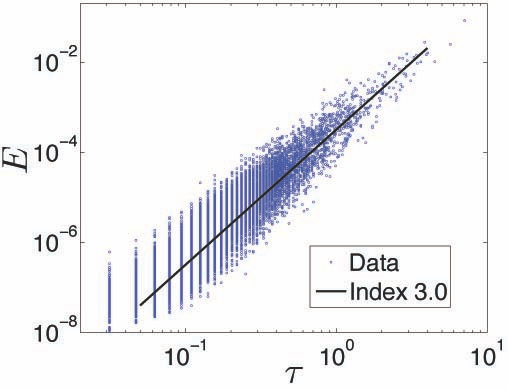}
   \centering
   \caption{\label{fig:scatter} Scatter plots of maximum length $L_\text{max}$, maximum width $W_\text{max}$, peak energy dissipation rate ${\cal E}_\text{max}$, peak volume $V_\text{max}$ (relative to the system volume), and dissipated energy $E$ versus the process duration $\tau$.}
 \end{figure}
We now describe the scaling relations between the various process characteristics. We show scatter plots of the different quantities versus process duration $\tau$ in Fig.~\ref{fig:scatter}. For clarity, these are only shown for Case 3, with similar scalings for all other cases. We find that $L_\text{max} \approx 6 W_\text{max} \approx 0.6 \tau$, ${\cal E}_\text{max} \sim V_\text{max} \sim \tau^2$, and $E \approx (3\times 10^{-4}) \tau^3$. We find that $T_\text{max}$ exhibits no evident correlation with $\tau$ or other quantities. Therefore, to a good approximation, the thickness is constant. These scalings are then consistent with the simple geometric estimates, $V_\text{max} \sim L_\text{max} W_\text{max} T_\text{max} \sim \tau^2$, ${\cal E}_\text{max} \sim V_\text{max} \eta j_\text{thr}^2 \sim \tau^2$, and $E = \int dt \int dV \eta j^2 \approx \tau V_\text{max} \eta j_\text{thr}^2 \sim \tau^3$ based on the other correlations, assuming that the thickness and typical current densities are constant.

A constraint between the indices of the distributions can be derived analytically if all processes are assumed to be single paths evolving with identical, rescaled functional forms. In this case, $E \sim {\cal E}_\text{max} \tau$ is an exact relation; see Appendix~\ref{appendix:constraint} for a derivation of this result. In addition, the distributions and scaling relations can be checked for self-consistency by using the conservation of probability. For example, one may suppose that $P({\cal E}_\text{max}) \sim {\cal E}_\text{max}^{-2}$, which implies that the distribution for peak energy dissipation rates is not dominated by weak or strong events. Then the measured scaling relations, ${\cal E}_\text{max} \sim \tau^2$ and $E \sim \tau^3 \sim {\cal E}_\text{max}^{3/2}$, fix the indices of the other distributions. In this case, $P(E)  = P({\cal E}_\text{max})\, {d{\cal E}_\text{max}/dE} \sim E^{-5/3}$, which is relatively close to our measured index near $-1.75$. The scalings then also imply that $P(\tau) \sim \tau^{-3}$, consistent with our measured index of $-3.2$. In general, it is clear that $P(E)$ should be somewhat shallower than $P({\cal E}_\text{max})$ due to integration across the duration, which increases with ${\cal E}_\text{max}$.

\subsection{Process evolution} \label{sec:evolve}

\begin{figure}
\includegraphics[width=8cm]{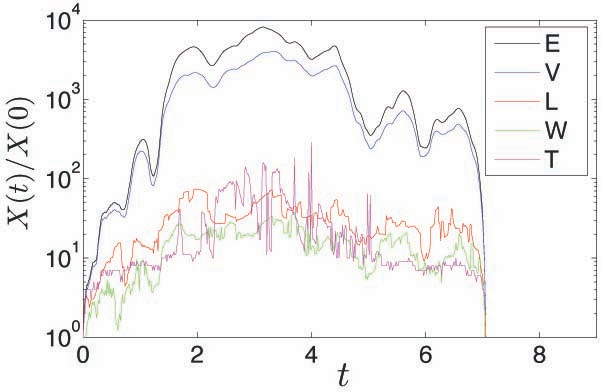}
\includegraphics[width=8cm]{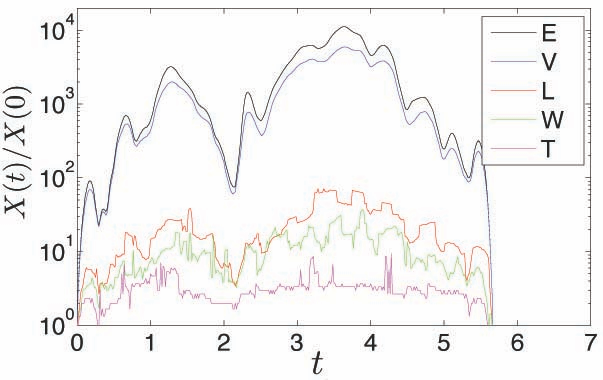}
   \centering
   \caption{\label{fig:history_sample} The evolution of several characteristics for the two longest processes for the $Re = 1250$ case. The curves correspond to energy dissipation rate ${\cal E}$ (black), volume $V$ (blue), length $L$ (red), width $W$ (green), and thickness $T$ (magenta).}
 \end{figure}

We now present results on the temporal evolution of individual processes. The following information is based on the time-series of instantaneous characteristics for each process, obtained from the constituent states at each snapshot. In general, the evolution of a given process is irregular and chaotic - in particular, long-lived processes are marked by frequent interactions and various phases of growth and decline. For example, the evolution of several characteristics for the two longest processes in Case 3 (at $j_\text{thr}/j_\text{rms} = 6.8$), which have durations of $\tau \approx 7.1$ and $5.7$, are shown in Fig.~\ref{fig:history_sample}. These processes begin by rapid growth, followed by a relatively steady phase that is randomly kicked via interactions, and end by rapid decay toward the threshold. We investigate the evolution of a typical process by averaging over all processes of a given duration.

\begin{figure}
\includegraphics[width=8cm]{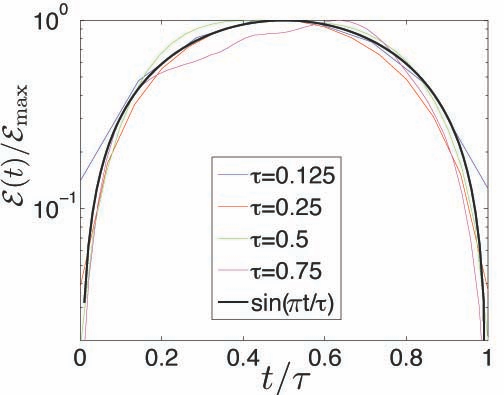}
\includegraphics[width=8cm]{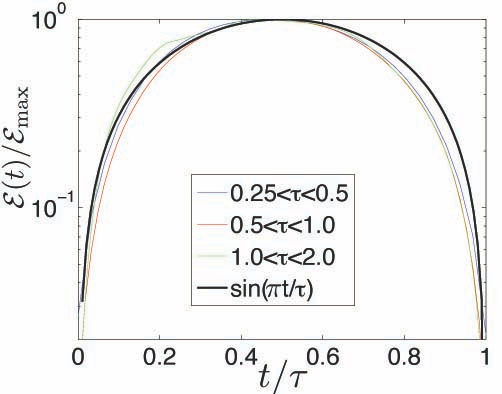}
   \centering
   \caption{\label{fig:hist_energy_average} Energy dissipation rate versus time, averaged for all processes of given durations (left) and all processes in given intervals of durations (right). Also shown in black is the fit by a sine function.}
 \end{figure}

To be concrete, we focus on the evolution of energy dissipation rate, ${\cal E}(t)$ for $0 < t < \tau$. Shown in the left panel of Fig.~\ref{fig:hist_energy_average} is the averaged energy dissipation rate normalized to peak energy dissipation rate, ${\cal E}(t/\tau)/{\cal E}_\text{max}$, versus time normalized to duration, $t/\tau$, for processes of durations $\tau \in \{ 0.125, 0.25, 0.5, 0.75 \}$ in Case 3. The evolution is well approximated by a single sine mode, ${\cal E}(t/\tau)/{\cal E}_\text{max} \approx \sin{(\pi t/\tau)}$, independent of $\tau$. Since the long-lived processes have a similar evolution as short-lived processes (with time normalized to duration and energy dissipation rate normalized to the corresponding peak value), it is reasonable to average the statistics over structures with varying $\tau$. This type of average is shown in the right panel of Fig.~\ref{fig:hist_energy_average} for all processes with $\tau$ in the intervals $\{(0.25,0.5), (0.5,1), (1,2)\}$. The averaged ${\cal E}(t/\tau)/{\cal E}_\text{max}$ continues to follow the same form up to $\tau \approx 2$, above which the statistical sample becomes limited.

\begin{figure}
\includegraphics[width=8cm]{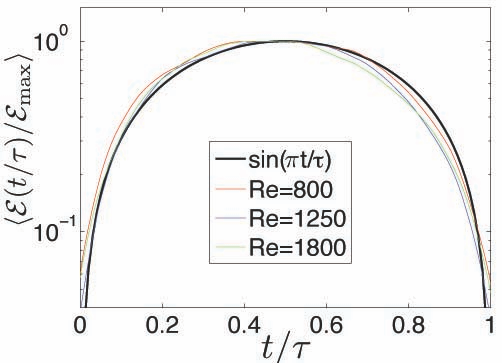}
\includegraphics[width=8cm]{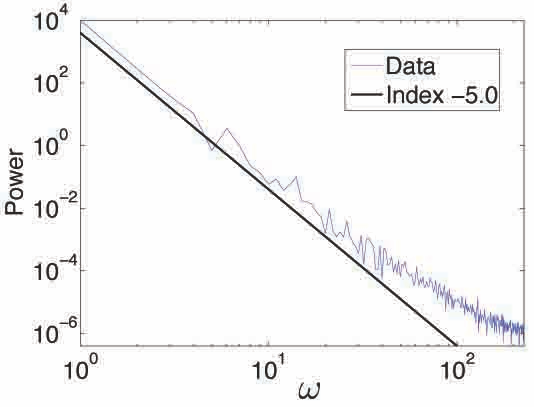}
   \centering
   \caption{\label{fig:hist_energy_superaverage} Left panel: the evolution of energy dissipation rate $\langle {\cal E}(t/\tau)/{\cal E}_\text{max} \rangle$ versus time $t/\tau$, with the average performed across processes of all durations. The fit by $\sin{(\pi t / \tau)}$ (shown in black) works very well. The colors correspond to $Re = 800$ (red), $Re = 1250$ (blue), and $Re = 1800$ (green). Right panel: the power spectrum of $\langle {\cal E}(t/\tau)/{\cal E}_\text{max} \rangle$ for the $Re = 1250$ case, showing a very steep descent as a power law with index near $-5.0$ at low $\omega$.}
 \end{figure}

We next perform an average over processes of \it all \rm durations to obtain $\langle{\cal E}(t/\tau)/{\cal E}_\text{max}\rangle$, shown for Cases 2-4 in Fig.~\ref{fig:hist_energy_superaverage}. It is clear that $\langle{\cal E}(t/\tau)/{\cal E}_\text{max}\rangle \approx \sin{(\pi t/\tau)}$ holds to a very good approximation. For a more quantitative analysis, we show the power spectrum of $\langle{\cal E}(t/\tau)/{\cal E}_\text{max}\rangle$ for Case 3 in the right panel of Fig.~\ref{fig:hist_energy_superaverage}, which has a very steep decline in power going approximately as $\omega^{-5}$ at low $\omega$, confirming that the $\omega = 1$ mode strongly dominates. The geometric characteristics $V$, $L$, $W$, and $T$ show a similar temporal evolution as ${\cal E}$, consistent with the strong correlations.

\begin{figure}
\includegraphics[width=8cm]{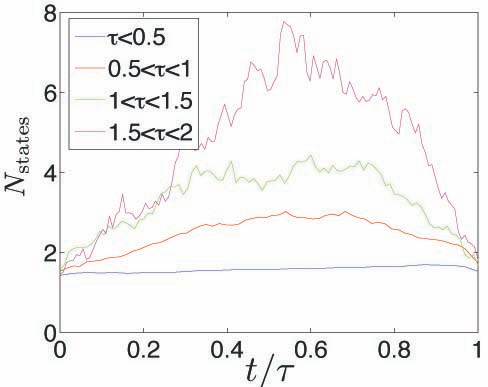}
\includegraphics[width=8cm]{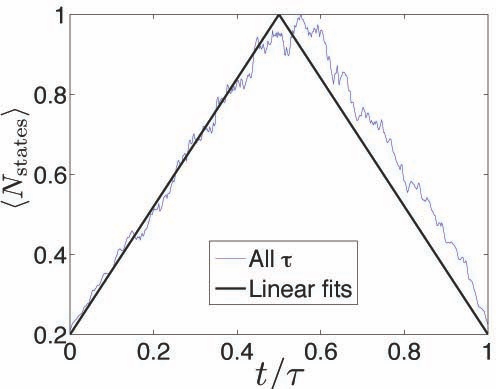}
   \centering
   \caption{\label{fig:hist_states} The instantaneous number of states in the process, $N_\text{states}(t/\tau)$, across the duration of the process, averaged for processes with durations in given intervals (left) and for processes with all durations (right).}
 \end{figure}

Next we consider the temporal evolution of the instantaneous number of states involved in the process, $N_\text{states}(t/\tau)$. This is shown for Case 3 in Fig.~\ref{fig:hist_states}, with averages performed across durations in the four intervals $\{(0,0.5),(0.5,1),(1,1.5),(1.5,2)\}$ (left panel) and across all durations (right panel). The functional form of $N_\text{states}(t)$ exhibits clear qualitative differences from ${\cal E}(t)$ and the geometric characteristics. The average across all durations can be approximated as a triangle function, i.e., linearly increasing in time and then linearly decreasing in time.

\begin{figure}
\includegraphics[width=8cm]{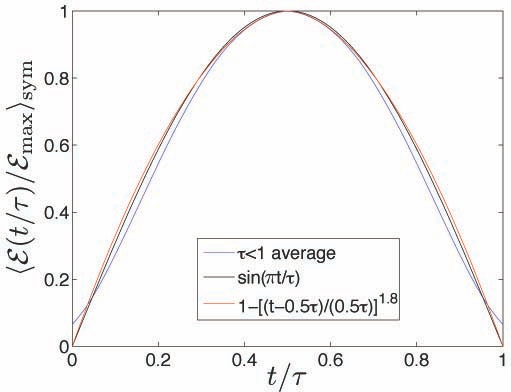}
\includegraphics[width=8cm]{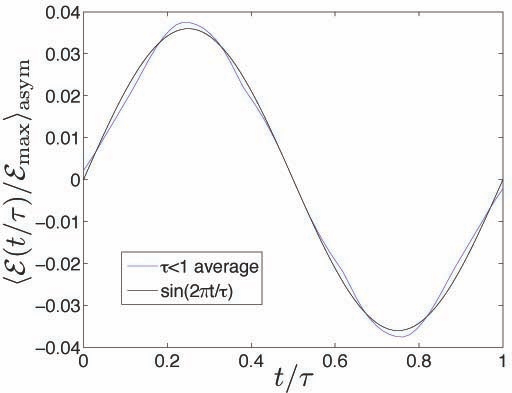}
   \centering
   \caption{\label{fig:hist_symmetry} Left panel: the symmetric part of $\langle {\cal E}(t/\tau)/{\cal E}_\text{max} \rangle$, averaged across processes with $\tau < 1$. The fit by $\sin{(\pi t / \tau)}$ (shown in black) and $1 - [(t-0.5\tau)/(0.5 \tau)]^{1.8}$ (shown in red) both work very well. Right panel: the corresponding antisymmetric part, fit by $0.036 \sin{(2 \pi t / \tau)}$ (shown in black).}
 \end{figure}

\begin{figure}
\includegraphics[width=8cm]{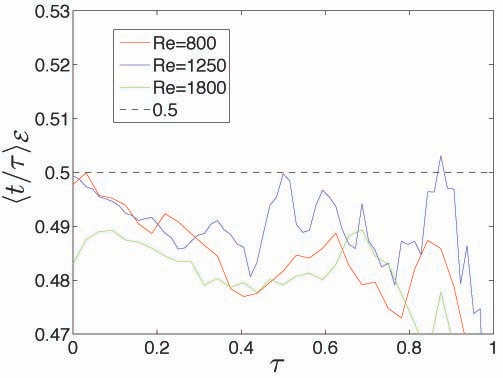}
\includegraphics[width=8cm]{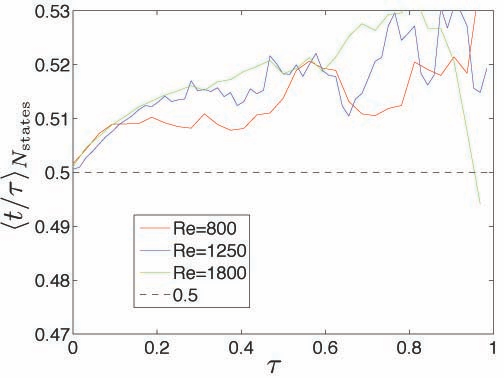}
   \centering
   \caption{\label{fig:hist_moment} The first moment, $\langle t/\tau \rangle$, of the evolution of energy dissipation rate ${\cal E}(t/\tau)/{\cal E}_\text{max}$ (left) and number of states $N_\text{states}(t/\tau)$ (right), versus process duration $\tau$ for $Re = 800$ (red), $Re = 1250$ (blue), and $Re = 1800$ (green). The curves are smoothed for clarity.}
 \end{figure}

Although ${\cal E}(t)$ and $N_\text{states}(t)$ are symmetric to a good approximation, a small asymmetric component can be discerned from Figs.~\ref{fig:hist_energy_superaverage} and \ref{fig:hist_states}. This asymmetry can be seen more clearly by decomposing the curve into symmetric and anti-symmetric parts,
\begin{align}
f_\text{sym}(t) &= \frac{f(t) + f(\tau-t)}{2} \nonumber \\
f_\text{asym}(t) &= \frac{f(t) - f(\tau-t)}{2} \, ,
\end{align}
where $f(t)$ is the given function on $0 < t < \tau$. Taking $f(t) = \langle{\cal E}(t/\tau)/{\cal E}_\text{max}\rangle$, we show the symmetric and anti-symmetric parts in Fig.~\ref{fig:hist_symmetry}. As noted before, the symmetric part is well fit by $\sin{(\pi t/\tau)}$. We also find that it can be equally well fit by $1 - [(t - 0.5\tau)/(0.5\tau)]^{1.8}$, which is nearly indistinguishable from the sine peak. We find that the anti-symmetric part of $\langle{\cal E}(t/\tau)/{\cal E}_\text{max}\rangle$ can be very well fit by $\sin{(2 \pi t/\tau)}$, with an amplitude of $0.036$. To investigate the asymmetry more precisely, we consider the first moments of the evolution curves,
\begin{eqnarray}
\langle t/\tau \rangle_f &=& \frac{\int_0^\tau (t/\tau) f(t) dt}{\int_0^\tau f(t) dt} \, ,
\end{eqnarray}
where deviation from $0.5$ is indicative of temporal asymmetry. We show $\langle t/\tau \rangle_{\cal E}$ and $\langle t/\tau \rangle_{N_\text{states}}$ for $\tau < 1$ in Fig.~\ref{fig:hist_moment}. We find that $\langle t/\tau \rangle_{\cal E}$ is very close to but slightly below $0.5$, while $\langle t/\tau \rangle_{N_\text{states}}$ is very close to but slightly above $0.5$. At small $\tau$, the displacement from $0.5$ initially grows with increasing $\tau$, but then asymptotes and becomes dominated by scatter at large $\tau$. Upon averaging over all durations, we find that for $Re = \{ 800,1250,1800 \}$, $\langle t/\tau \rangle_{\cal E} = \{0.483,0.483,0.476\}$ while $\langle t/\tau \rangle_{N_\text{states}}= \{0.517,0.517,0.522\}$. Incidentally, the degree of asymmetry is comparable for both types of measurements, although in opposite directions.

\section{Discussion} \label{sec6}

\subsection{Comparison to solar flare observations}

In this subsection, we compare our statistical results for dissipative processes in MHD turbulence with the statistical properties of solar flares, taken from a number of observational studies. This comparison is not intended to draw any direct conclusions about solar flares, since our simulations are physically inadequate for describing the overall dynamics of the solar corona, although they may describe the turbulence that develops at small scales. Instead, the present comparison is motivated by the fact that solar flares are the best-observed natural example of intermittent energy dissipation in large-scale magnetized plasmas. For a more direct comparison, simulations of line-tied MHD \citep{galsgaard_nordlund_1997, ng_bhattacharjee_1998, ng_bhattacharjee_2012, wan_etal_2014} or other numerical models of the corona \citep{bingert_peter_2011, bingert_peter_2013} may be investigated.

The properties of solar flares are obtained from observations of extreme UV (EUV), soft X-ray, and hard X-ray emissions by applying a methodology similar to the one used in this paper. However, there are several important, unavoidable methodological differences that may affect the comparison. First, the emission may not be in direct association with the dissipation, making it nontrivial to infer the dissipation from the spectral amplitude. Indeed, although hard X-rays are thought to be promptly powered by dissipative magnetic reconnection events, soft X-rays and EUV can originate from aftereffects including chromospheric evaporation and cooling. Second, images of solar flares are projected onto a 2D plane, reducing the available information. In addition, there are several physical differences between driven, incompressible MHD turbulence and the solar corona. In contrast to volumetrically-driven turbulence in a periodic box, flares in the solar corona are generally modeled by force-free MHD with slowly-driven, line-tied boundary conditions. Additional plasma physics arising from kinetic and two-fluid effects may be important during magnetic reconnection. Following the reconnection event, other physical processes including chromospheric evaporation, radiative cooling, and thermal conduction may affect the decay of the solar flare. Nevertheless, we proceed with the comparison.

In order to make a tangible comparison, we focus on a handful of studies which are methodologically most similar to our present work. These are the papers by \cite{uritsky_etal_2007} (U07), \cite{uritsky_etal2013} (U13), and \cite{aschwanden_etal_2014} (A14). These studies take extreme UV images of the corona and magnetograms of the photosphere to identify 3D spatiotemporal dissipative processes. The range of indices for the distributions and scalings in these studies, along with our results, are shown in Table~\ref{table-corona}. The statistics for dissipated energy and length agree favorably with our results, whereas they appear to differ for durations. The sub-diffusive growth of solar flares \citep{aschwanden_2012b, aschwanden_etal_2013}, which has been modeled in the framework of self-organized criticality \citep{aschwanden_2012a}, also appears to be at odds with the evolution of processes measured in our work.

\begin{table*}[h!b!p!]
\caption{Comparison of distributions and scalings with solar flare studies \newline}
\begin{tabular}{|c|c|c|c|c|}
	\hline
\hspace{0.5 mm} Quantity \hspace{0.5 mm}  & \hspace{0.2 mm} MHD turbulence & \hspace{0.2 mm} U07 \hspace{0.2 mm}  & \hspace{0.2 mm} U13 \hspace{0.2 mm} & \hspace{0.2 mm} A14 \hspace{0.2 mm} \\
	\hline
Index for $P(E)$ & $1.75$ & $1.6-1.7$ & $1.5$ & $1.8-2.2$  \\
Index for $P({\cal E}_\text{max})$ & 2.0 & $-$ & $-$ & $2.1-2.5$ \\
Index for $P(L)$ & $3.2$ & $-$ & $2.5-2.9$& $3.5-4.1$ \\
Index for $P(\tau)$ & $3.2$ & $1.9-2.1$ & $2.0-2.2$ & $2.2-2.6$ \\
$\log{E} / \log{L}$ & $3.0$ & $3.0-3.6$ & $3.0-3.1$ & $2.5-2.6$ \\
$\log{\tau}/\log{L}$ & $1.0$ & $1.8-2.3$ & $1.2-1.4$ & $-$ \\
	\hline
\end{tabular}
\centering
\label{table-corona}
\end{table*}

Comparing to more general studies of solar flares, our distribution for dissipated energy, with index near $-1.75\pm0.1$, is close to the analogous measurements for total energy released in solar flares identified from hard X-rays, generally having an index quoted to be between $-1.7$ and $-1.8$ \citep[e.g.,][]{aschwanden_etal_2000b, bromund_etal_1995, christe_etal_2008}. Similarly, our distribution for peak energy dissipation rate ${\cal E}_\text{max}$ with index $-2.0 \pm 0.1$ is close to observations of peak hard X-ray flux \citep[e.g.,][]{bromund_etal_1995} and soft X-ray flux \citep[e.g.,][]{aschwanden_freeland_2012}. Our distribution for duration with index near $-3.2 \pm 0.2$ is somewhat steeper than the indices ranging between $-2.2$ and $-3.0$ for solar flare durations \citep{crosby_etal1993, bromund_etal_1995, veronig_etal_2002}, although it is closer to the index for rise times, given as $-3.4$ in \cite{christe_etal_2008} and $-3.2$ (during solar maximum; shallower otherwise) in \cite{aschwanden_freeland_2012}. One final point of comparison is the asymmetry of processes, recalling that in our case, a process tends to dissipate slightly more energy at early times than late times. This is qualitatively in agreement with observations of solar and stellar flares, although the asymmetry appears to be much more pronounced in flares. The asymmetry can be defined from the rise time $t_\text{rise}$ and decay time $t_\text{decay}$ as $A_\text{ev} = (t_\text{decay} - t_\text{rise})/(t_\text{decay} + t_\text{rise})$; this is found to be $0.2$ for X-ray flares, giving a peak at approximately $40\%$ of the flare duration \citep{christe_etal_2008}. In contrast, the asymmetry of processes in our simulations, based on this definition, is $0.034$.

In summary, the energetic and geometric statistical properties of dissipative events in MHD turbulence are consistent with solar flare observations, whereas the durations and temporal asymmetries present a noticeable discrepancy. The differences may be due to the Neupert effect, in which the chromospheric evaporation prolongs the decay of a flare observed in soft X-rays relative to hard X-rays \citep{neupert_1968, dennis_zarro_1993}. This would explain why the distribution of process durations in our present work matches the distribution of solar flare rise times better than their total durations.

The nontrivial similarities in the statistical properties between dissipative events in MHD turbulence and in the solar corona leaves open the possibility that MHD turbulence plays a governing role in the intermittency of the coronal energetics. This possibility has been advocated in numerous past studies \citep[e.g.,][]{georgoulis_2005, uritsky_etal_2007, uritsky_etal2013} as an alternative to self-organized criticality. A more careful three-way comparison of the temporal statistics of dissipative events in MHD turbulence, self-organized criticality, and observations of the solar corona is left for future consideration.

\subsection{Implications for MHD turbulence}

The methodology developed in this paper is complementary to the tools conventionally used to probe turbulence and its intermittency. It also has several advantages over the other methods, despite the relatively complex numerical implementation and the necessity of a large data set. For example, given a relatively meager $Re = 800$, the distribution for dissipated energy per process shows a power law across nearly three decades in energy, whereas the inertial range is barely discernable in the corresponding energy spectrum at the same $Re$. This large separation of scales may be attributed to the additional information given by the temporal dynamics of the turbulence. The analysis of structures also naturally describes the anisotropy and inhomogeneity of the dynamics, which is often challenging for other methods.

In this work and in our previous study of spatial structures \citep{zhdankin_etal2014}, we found that the distribution for the energy dissipation rates of states has an index close to the critical index of $-2$. This suggests that, at any given time, structures of all energy dissipation rates (in the inertial range) contribute almost equally to the overall energy dissipation rate. This may be a manifestation of the scale-invariance of inertial-range turbulence, since a distribution for energy dissipation rates with the critical index is equivalent to the energy dissipation being evenly spread across structures of all lengths \citep{zhdankin_etal2014}.

Our present work goes beyond this previous result by establishing that the dissipated energy in evolving structures (processes) has a power law distribution with index shallower than the critical index, namely, with an index near $-1.75 \pm 0.1$. This implies that intense dissipative events, i.e., large-scale and long-lived coherent structures, dominate the overall energy dissipation. This is a consequence of the linear scaling of duration with maximum length, which causes the distribution of dissipated energy to be shallower than the distribution of energy dissipation rates.

The distributions, scalings, and evolution of processes appear to be insensitive to the $Re$ sampled in our simulations. This suggests that these statistics may be valid for asymptotically large $Re$, relevant for space and astrophysical turbulence. This also implies that the results may be universal, i.e., insensitive to the mechanisms of energy input and dissipation, although this should be verified in the future by varying the boundary conditions, forcing mechanisms, and dissipation mechanisms. Examples of astrophysically-relevant driving mechanisms include the magnetorotational instability for accretion disks \citep{balbus_hawley_1991} and line-tied driving for the force-free solar corona. It is also possible for the nature of intermittency to undergo a transition at sufficiently large $Re$, due to instabilities for large and morphologically complex structures. Therefore, it is important to verify our results in future simulations of MHD turbulence with larger $Re$, where more precise power-law fits can be obtained and a systematic study of the $Re$ dependence can be investigated. It is challenging to apply our methodology, in its present form, to direct numerical simulations with larger $Re$, since both the spatial resolution and time cadence must be increased, making it impractical to store the full sequence of data snapshots. It may be necessary to perform the bulk of the analysis in parallel with the simulations, rather than analyzing post-processed snapshots as was done in this work. Alternatively, the amount of information used for the analysis may be reduced by, e.g., filtering out large-scale or small-scale modes. This is left for future consideration.

Our methodology provides a new avenue to investigating temporal asymmetry, which was previously inferred in studies of MHD turbulence through the third-order moment or rate of energy flux \citep[e.g.,][]{macbride_etal_2008, podesta_2008, wan_etal_2010b} and field-line diffusion \citep{beresnyak_2014}. The temporal asymmetry in our case is measured in the larger number of divisions than mergers, the tendency of the number of states in a process to be larger at late times than early times, and the tendency of the energy dissipation rate and geometric characteristics to be larger at early times than late times. Temporal asymmetry may occur from the onset of an instability, such as the tearing instability in resistive MHD or avalanches in critically self-organized systems. However, the asymmetry measured in our simulations is relatively small and does not significantly increase for larger structures that may cross an instability threshold. Furthermore, the processes do not exhibit the impulsiveness expected when an instability is triggered. Therefore, we find it unlikely that the instability of structures plays a role in our simulations, although it will be an important signature to search for in future studies. Indeed, the tearing instability is expected to occur for laminar, 2D current sheets when they become sufficiently thin, which may occur at $Rm \sim 10^4$ \citep[e.g.,][]{ loureiro2007, bhattacharjee2009, uzdensky2010}. It is conceivable that current sheets in a turbulent medium become unstable at different (possibly lower) $Rm$ than naively expected \citep{loureiro_etal2009}; it is also possible that the instability is entirely absent.

In our case, the temporal asymmetry may be linked to the turbulent energy cascade. Specifically, the inertial range of 3D MHD turbulence is characterized by a direct energy cascade from large scales to small scales. Therefore, turbulent eddies cause large structures with inertial-range lengths and widths to be broken into smaller structures, leading to a surplus of divisions over mergers, as well as more states at late times in a process. This can also explain why the energy dissipation rate and geometric characteristics are larger at early times, since a single large state may accomodate a higher current density than many individual states. Since the dynamics are otherwise time-symmetric in the inertial range, this asymmetry can be relatively weak. We note that another distinct contribution to the asymmetry can be from the dissipative term directly (rather than the cascade through the inertial range), relevant for states with lengths and widths that are near the dissipation scale. It is left to future work to better quantify the asymmetry and its origins, and to relate the measured quantities to the energy cascade rate.

\section{Conclusion} \label{sec7}

The methodology developed in this paper, partly inspired by similar approaches in observational astrophysics and self-organized criticality, presents a lucid picture of intermittency in MHD turbulence with dissipative structures playing a central role. Previously, this picture was incomplete, based on a fraction of the available information. By exploring the temporal dimension, we completed this picture and found a richer, more valuable perspective of turbulence. 

This work demonstrates that the statistical analysis of spatiotemporal dissipative structures, i.e., processes or flares, can lead to concrete physical insights for intermittency in turbulence. The methodology is a natural but nontrivial extension of the methodology applied to characterize spatial dissipative structures in previous studies. We applied this analysis to numerical simulations of MHD turbulence to arrive at the following basic conclusions, which are insensitive to the numerical parameters (cadence, resolution, and threshold). We found that resistive energy dissipation occurs in current sheets that participate in intense, complex, long-lived processes with durations that may span several large eddy turnover times. These processes are analogous to flares in the solar corona and other astrophysical systems. The durations of the processes are directly proportional to the maximum lengths, providing a strong link between the spatial and temporal behavior. The energy dissipated in these intense processes is distributed as a power law with index near~$-1.75$, implying the dominance of large, intense flares. Incidentally, this index is consistent with observed energy distributions of solar flares. Processes are weakly asymmetric in time, dividing more often than merging. The averaged temporal evolution for the energy dissipation rate (and geometric properties) of the processes exhibits a nearly time-symmetric, sine-like form that is applicable to processes of all durations that were robustly sampled in our study.

The insights gained from this methodology may help to understand magnetic reconnection, heating, and particle acceleration in space and astrophysical plasmas. Eventually, this methodology may also help narrow the imposing gap between our theoretical knowledge of energy dissipation in magnetized plasmas and the observational problem of coronal heating, and to uncover the similarities and differences between self-organized criticality and turbulence. In the future, we hope that the additional utilization of the methods described in this work will further evolve our present picture of turbulence. Indeed, the picture presented in this paper is incomplete in that it sidesteps the intermittency of vorticity and viscous dissipation, as well as its coupling to the current density and resistive dissipation, which should be scrutinized in a later work. In addition, we believe that further phenomenological modeling of the spatiotemporal dissipative structures in MHD turbulence is a promising pursuit.

\acknowledgements

The authors would like to thank Jean Carlos Perez for his support in conducting the numerical simulations. This work was supported by NASA grant No. NNX11AE12G, US DOE grants DE-SC0003888, DE-SC0008409, DE-SC0008655, and NSF Grants AST-1411879 and NSF PHY11-25915. SB appreciates the hospitality and support of the Kavli Institute for Theoretical Physics, University of California, Santa Barbara, where part of this work was conducted.

\newpage

\appendix
\section{Appendix}

\subsection{Tables of occurrence rates} \label{appendix:tables}

The tables in this Appendix show the occurrence rates used to compute ratios in Tables \ref{table-cases}, \ref{table-cadence}, and \ref{table-threshold}. Table~\ref{table-cases2} compares all cases, Tables~\ref{table-cadence2} and \ref{table-cadence3} show varying cadence for Case 1 and Case 3, and Table~\ref{table-threshold2} shows varying threshold for Case 3.

\begin{table*}[h!b!p!]
\caption{Aggregate quantities in four simulations ($j_\text{thr}/j_\text{rms} \approx 6.8$, $\Delta{t} = 1/32$) \newline}
\begin{tabular}{|c|c|c|c|c|} 
	\hline
\hspace{0.5 mm} Quantity (per $\tau_\text{eddy}$) \hspace{0.5 mm}  & \hspace{2 mm}$256^3, Re = 800$\hspace{2 mm}    &   \hspace{2 mm}$512^3, Re = 800$\hspace{2 mm} &\hspace{2 mm}$512^3, Re = 1250$  & \hspace{2 mm}$512^3, Re = 1800$\hspace{2 mm}  \hspace{2 mm} \\
	\hline
$N_\text{path}$ & 4202 & 5295 & 15392 & 34275 \\
$N_\text{proc}$ & 914 & 1271 & 4272 & 11608 \\
$N_\text{isol}$ & 715 & 956  & 3600 & 10416 \\
$N_\text{int}$ & 199 & 315  & 672 & 1193 \\
$N_\text{div}$ & 1494 & 1663  & 5352 & 11585 \\
$N_\text{mer}$ & 1253 & 1290  & 4276 & 9540 \\ 
$N_\text{des}$ & 1746 & 2458  & 7088 & 17312 \\
$N_\text{form}$ & 1520 & 2096 & 6283 & 15766 \\ 
$N_\text{3-vert}$ & 1449 & 1527 & 3886 & 7745  \\
$N_\text{n-vert}$ & 494 & 539 & 1884 & 4051  \\ 
$\langle N_\text{state} \rangle$ & 194 & 288 & 657 & 1328  \\
	\hline
\end{tabular}
\centering
\label{table-cases2} 
\end{table*}

\begin{table*}[h!b!p!]
\caption{Aggregate quantities with changing cadence (Case 1: $256^3, Re=800, j_\text{thr}/j_\text{rms} \approx 6.8$) \newline}
\begin{tabular}{|c|c|c|c|c|c|} 
	\hline
\hspace{0.5 mm} Quantity \hspace{0.5 mm}  & \hspace{0.2 mm} $\Delta{t} = 1/64$ \hspace{0.2 mm}  & \hspace{0.2 mm}  $\Delta{t} = 1/32$ \hspace{0.2 mm} & \hspace{0.2 mm} $\Delta{t} = 1/16$ \hspace{0.2 mm} & \hspace{0.2 mm} $\Delta{t} = 1/8$ \hspace{0.2 mm}   & \hspace{0.2 mm} $\Delta{t} = 1/4$ \hspace{0.2 mm} \\
	\hline
$N_\text{path}$ & 6852 & 4202 & 2426 & 1360 & 720 \\
$N_\text{proc}$ & 1136 & 914 & 959 & 921 & 629 \\
$N_\text{isol}$ & 894 & 715 & 834 & 867 & 613 \\
$N_\text{int}$ & 241 & 199 & 126 & 54 & 16 \\
$N_\text{div}$ & 2614 & 1494 & 666 & 199 & 39 \\
$N_\text{mer}$ & 2277 & 1253 & 514 & 141 & 27 \\ 
$N_\text{des}$ & 2295 & 1746 & 1418 & 1078 & 663 \\ 
$N_\text{form}$ & 1984 & 1520 & 1276 & 1022 & 651 \\ 
$N_\text{3-vert}$ & 2953 & 1449 & 512 & 125 & 24 \\ 
$N_\text{n-vert}$ & 775 & 494 & 235 & 71 & 14 \\ 
$\langle N_\text{state} \rangle$ & 194 & 194 & 193 & 192 & 189 \\ 
	\hline
\end{tabular}
\centering
\label{table-cadence2}
\end{table*}

\begin{table*}[h!b!p!]
\caption{Aggregate quantities with changing cadence (Case 3: $512^3, Re=1250, j_\text{thr}/j_\text{rms} \approx 6.8$) \newline}
\begin{tabular}{|c|c|c|c|c|c|} 
	\hline
\hspace{0.5 mm} Quantity \hspace{0.5 mm}  & \hspace{0.2 mm} $\Delta{t} = 1/64$ \hspace{0.2 mm}  & \hspace{0.2 mm}  $\Delta{t} = 1/32$ \hspace{0.2 mm} & \hspace{0.2 mm} $\Delta{t} = 1/16$ \hspace{0.2 mm} & \hspace{0.2 mm} $\Delta{t} = 1/8$ \hspace{0.2 mm}   & \hspace{0.2 mm} $\Delta{t} = 1/4$ \hspace{0.2 mm} \\
	\hline
$N_\text{path}$ & 27500 & 15390 & 8731 & 4748 & 2454 \\
$N_\text{proc}$ & 3311 & 4272 & 4908 & 3704 & 2197 \\
$N_\text{isol}$ & 2316 & 3600 & 4600 & 3595 & 2163 \\
$N_\text{int}$ & 994 & 672 & 308 & 109 & 34 \\
$N_\text{div}$ & 11900 & 5352 & 1816 & 495 & 116 \\
$N_\text{mer}$ & 10060 & 4276 & 1447 & 382 & 90 \\ 
$N_\text{des}$ & 8364 & 7088 & 5964 & 4031 & 2283 \\ 
$N_\text{form}$ & 6954 & 6283 & 5685 & 3942 & 2264 \\ 
$N_\text{3-vert}$ & 10750 & 3886 & 1099 & 261 & 57 \\ 
$N_\text{n-vert}$ & 3949 & 1884 & 667 & 182 & 44 \\ 
	\hline
\end{tabular}
\centering
\label{table-cadence3} 
\end{table*}

\begin{table*}[h!b!p!]
\caption{Aggregate quantities with changing threshold (Case 3: $512^3, Re = 1250$) \newline}
\begin{tabular}{|c|c|c|c|c|}
	\hline
\hspace{0.5 mm} Quantity \hspace{0.5 mm}  & \hspace{0.2 mm} $j_\text{thr}/j_\text{rms} \approx 9.6$ \hspace{0.2 mm}  & \hspace{0.2 mm}  $j_\text{thr}/j_\text{rms} \approx 8.2$ \hspace{0.2 mm} & \hspace{0.2 mm} $j_\text{thr}/j_\text{rms} \approx 6.8$ \hspace{0.2 mm} & \hspace{0.2 mm} $j_\text{thr}/j_\text{rms} \approx 5.5$ \hspace{0.2 mm} \\
	\hline
$N_\text{path}$ & 7975 & 14270 & 27500 & 54630  \\
$N_\text{proc}$ & 1105 & 1777 & 3311 & 6314  \\
$N_\text{isol}$ & 694 & 1200 & 2316 & 4565  \\
$N_\text{int}$ & 341 & 577 & 994 & 1749  \\
$N_\text{div}$ & 3077 & 5757 & 11900 & 26390  \\
$N_\text{mer}$ & 2539 & 4788 & 10060 & 22785 \\ 
$N_\text{des}$ & 2694 & 4705 & 8364  & 12540 \\ 
$N_\text{form}$ & 2210 & 3858 & 6954 & 10900 \\ 
$N_\text{3-vert}$ & 3081 & 5557 & 10750 & 22120 \\ 
$N_\text{n-vert}$ & 964 & 1837 & 3949 & 8854 \\ 
$\langle N_\text{state} \rangle$ & 190 & 343 & 657 & 1287  \\ 
	\hline
\end{tabular}
\centering
\label{table-threshold2}
\end{table*}

\subsection{Constraints between indices of distributions} \label{appendix:constraint}

If we suppose that all processes are single paths that evolve with identical (but rescaled) functional forms for the energy dissipation rate, then we can derive nontrivial constraints between the indices of the various distributions given in Sec.~\ref{sec:prob}. Specifically, for simplicity, assume that all processes of duration $\tau$ consist of a single path with the energy dissipation rate ${\cal E}_\tau(t) = {\cal E}_{\text{max}}(\tau) f(t/\tau)$ for $0 < t < \tau$, where the universal shape function $f(x)$ satisfies $f(0) = f(1) = 0$, $0 \le f(x) \le 1$ and $\sup{\{f(x)\}} = 1$ for $0 < x < 1$. If we assume power-law distributions for all quantities, then they are given by
\begin{align}
P(E) \sim E^{-\alpha} \nonumber \\
P({\cal E}) \sim {\cal E}^{-\beta} \nonumber \\
P({\cal E}_\text{max}) \sim {\cal E}_\text{max}^{-\beta} \nonumber \\
P(\tau) \sim \tau^{-\gamma} \, ,
\end{align}
with the constraint
\begin{align}
(\beta - 1)(\gamma - \alpha) = (\gamma - 1) (\alpha - 1) \, .
\end{align}
The corresponding scaling relations are given by
\begin{align}
E &\sim \tau^{(\gamma - 1)/(\alpha - 1)} \nonumber \\
{\cal E}_\text{max} &\sim \tau^{(\gamma - \alpha)/(\alpha - 1)} \, .
\end{align}
Normalizable distributions require $\gamma > \alpha$. For the measured value of $\beta = 2$, $\gamma = 1/(2 - \alpha)$ is required, so that $1 < \alpha < 2$. An example set of parameters satisfying this constraint are $\beta = 2$, $\gamma = 3$, and $\alpha = 5/3$, along with scalings $E \sim \tau^3$ and ${\cal E}_\text{max} \sim \tau^2$, all of which are consistent with the results in this paper.

The derivation is as follows. We first relate the distribution of instantaneous energy dissipation rates, $P({\cal E})$, measured from states at random, to the distribution of peak energy dissipation rates, $P({\cal E}_\text{max})$, measured from processes. Assuming that one samples a random value ${\cal E}$ from ${\cal E}_\tau(t)={\cal E}_{\text{max}}(\tau) f(t/\tau)$ with uniform time sampling, so $P(t) = 1/\tau$, we obtain the distribution of energy dissipation rates from a process of duration $\tau$,
\begin{align}
P({\cal E}|\tau) &= \left| \frac{dt}{d{\cal E}} \right| P(t) = \frac{1}{{\cal E}_{\text{max}}(\tau)} \sum_{i = 1}^n{\left| \frac{dx}{df(x)} \right|_{x = x_i}} \equiv \frac{1}{{\cal E}_{\text{max}}(\tau)} g\left(\frac{{\cal E}_\text{max}(\tau)}{{\cal E}}\right) \, ,
\end{align}
where $x_i$ ($i = 1,\dots,n$) are the $n$ roots of $f(x_i) - {\cal E}/{\cal E}_{\text{max}}(\tau)$, and we have defined the function $g(y)$. The total distribution of energy dissipation rates is then
\begin{align}
P({\cal E}) &= \int^\infty_{\tau_\text{min}} d\tau P(\tau) P({\cal E}|\tau) \nonumber \\ 
&= \int^\infty_{\tau_\text{min}} d\tau P(\tau) \frac{1}{{\cal E}_{\text{max}}(\tau)} g\left(\frac{{\cal E}_\text{max}(\tau)}{{\cal E}}\right) \nonumber \\
&= \int^\infty_{\cal E} d{\cal E}_\text{max} \frac{P({\cal E}_\text{max})}{{\cal E}_{\text{max}}} g\left(\frac{{\cal E}_\text{max}}{{\cal E}}\right) \, ,
\label{magic}
\end{align}
where $\tau_\text{min}$ is defined such that ${\cal E}_{\text{max}}(\tau_\text{min}) = {\cal E}$. The lower bound of the integral is required since processes with durations $\tau < \tau_\text{min}$ do not reach high enough energy dissipation rates to contribute to the distribution. Now assume that $P({\cal E}_\text{max}) \sim {\cal E}_\text{max}^{-\beta}$. Then Eq.~\ref{magic} becomes
\begin{align}
P({\cal E}) &\sim \int^\infty_{\cal E} d{\cal E}_\text{max} {\cal E}_{\text{max}}^{-\beta-1} g\left(\frac{{\cal E}_\text{max}}{{\cal E}}\right)  \nonumber \\
&\sim {\cal E}^{-\beta} \int^\infty_1 dy y^{-\beta-1} g(y) \nonumber \\
&\sim {\cal E}^{-\beta} , \label{magic2}
\end{align}
where $y = {\cal E}_\text{max}/{\cal E}$. Therefore, assuming the integral in Eq.~\ref{magic2} converges, $P({\cal E})$ has the same index as $P({\cal E}_\text{max})$. To relate this to other indices, note that the dissipated energy $E(\tau)$ per process is given by
\begin{align}
E(\tau) &= \int_0^\tau dt {\cal E}_\tau(t) = {\cal E}_{\text{max}}(\tau) \tau \int_0^1 dx f\left(x\right) \sim  {\cal E}_{\text{max}}(\tau) \tau \, ,
\end{align}
where $x = t/\tau$, and the integral $\int_0^1 dx f\left(x\right)$ evaluates to a constant of order unity; hence, $E \sim {\cal E}_\text{max} \tau$ is exactly satisfied. Assuming $P(\tau) \sim \tau^{-\gamma}$ and $P(E) \sim E^{-\alpha} $, we can find the exponent $\lambda$ for $E \sim \tau^\lambda$,
\begin{align}
\frac{dE}{d\tau} = \frac{P(\tau)}{P(E)} &\implies \tau^{\lambda-1} = \frac{\tau^{-\gamma}}{\tau^{-\lambda \alpha}} \implies \lambda = \frac{\gamma - 1}{\alpha - 1} \, .
\end{align}
Hence, ${\cal E}_\text{max} \sim E/\tau \sim \tau^{-1+(\gamma-1)/(\alpha-1)} \sim \tau^{(\gamma-\alpha)/(\alpha-1)}$ and
\begin{align}
P({\cal E}_\text{max}) \sim \frac{d \tau}{d {\cal E}_\text{max}} P(\tau) \sim \tau^{1 - (\gamma-\alpha)/(\alpha - 1) -\gamma} \sim {\cal E}_\text{max}^{-1-(\gamma-1)(\alpha-1)/(\gamma - \alpha)} \, .
\end{align}
Therefore we have $\beta = 1 + (\gamma - 1)(\alpha - 1)/(\gamma - \alpha)$, as required.


\end{document}